\documentclass[reqno,12pt]{article}
\usepackage[T1]{fontenc}
\usepackage[ansinew]{inputenc}
\usepackage{amsmath}
\usepackage{amssymb}
\usepackage{graphicx}
\usepackage{xcolor}
\usepackage{color}
\definecolor{darkblue}{cmyk}{0.9,0.9,0,0}
\usepackage[colorlinks=true,linkcolor=darkblue,citecolor=darkblue,urlcolor=darkblue]{hyperref}
\usepackage{epsfig}
\usepackage{mdframed}
\usepackage{cite}

\usepackage{mathrsfs}

\newcommand{\bunderbrace}[2]{%
  \begin{array}[t]{@{}c@{}}
  \underbrace{#1}\\
  #2
  \end{array}
}

\newcommand*\dif{\mathop{}\!\mathrm{d}}
\newcommand{\normord}[1]{:\mathrel{#1}:}

\newcommand{\beq}{\begin{equation}}
\newcommand{\eeq}{\end{equation}}
\newcommand\beqa{\begin{eqnarray}}
\newcommand\eeqa{\end{eqnarray}}
\newcommand\bea{\begin{array}}
\newcommand\eea{\end{array}}

\newcommand\RE{{\rm Re}\,}

\def\XXint#1#2#3{{\setbox0=\hbox{$#1{#2#3}{\int}$}
\vcenter{\hbox{$#2#3$}}\kern-.5\wd0}}

\newcommand{\neqa}{\nonumber\end{eqnarray}}
\newcommand{\la}[1]{\label{#1}}

\newcommand{\Tr}{{\rm Tr}}

\renewcommand{\d}{\partial}

\newcommand{\<}{{\langle}}
\renewcommand{\>}{{\rangle}}

\newcommand{\re}{\relax{\rm I\kern-.18em R}}

\renewcommand{\sp}{p\hspace{-.40em}/}

\def\su2{{SU(2)}}

\def\a{{\alpha}}

\def\[{\left[}
\def\]{\right]}

\def\a{\alpha}

\def\({\left(}
\def\){\right)}
\def\[{\left[}
\def\]{\right]}

\def\<{\langle}
\def\>{\rangle}

\def\i2{\frac{i}{2}}

\def\spi{\relax{\rm \pi\kern-0.5em /}}
\def\sA{\relax{\rm A\kern-0.5em /}}
\def\sp{\relax{\rm p\kern-0.5em /}}
\def\sd{\relax{\rm \d\kern-0.5em /}}
\def\sk{\relax{\rm k\kern-0.5em /}}
\def\sn{\relax{\rm n\kern-0.5em /}}
\def\sl{\relax{\rm l\kern-0.5em /}}
\def\sP{\relax{\rm P\kern-0.7em /}}
\def\sBethe{\relax{\rm \Bethe\kern-0.5em /}}
\def\cN{{\cal N}}

\usepackage{varioref}
\usepackage{makeidx}
\makeindex

\mdfdefinestyle{MyFrame}{%
    linecolor=black,
    outerlinewidth=2pt,
    innertopmargin=4pt,
    innerbottommargin=4pt,
    innerrightmargin=4pt,
    innerleftmargin=4pt,
        leftmargin = 4pt,
        rightmargin = 4pt
        }
        
\usepackage[english]{babel}
\begin{document}


\thispagestyle{empty}

\renewcommand{\thefootnote}{\fnsymbol{footnote}}
\setcounter{page}{1}
\setcounter{footnote}{0}
\setcounter{figure}{0}

\vspace{-0.4in}

\hfill CALT-TH 2024-033

\begin{center}
$$$$
{\Large\textbf{\mathversion{bold}
Light-Ray Wave Functions and Integrability
}\par}
\vspace{1.0cm}

\textrm{Alexandre Homrich${}^\text{\tiny 1,2}$, David Simmons-Duffin${}^\text{\tiny 3}$ and Pedro Vieira${}^\text{\tiny 1,4}$}
\\ \vspace{1.2cm}
\footnotesize{\textit{
$^\text{\tiny 1}$Perimeter Institute for Theoretical Physics,
Waterloo, Ontario N2L 2Y5, Canada  \\
$^\text{\tiny 2}$Kavli Institute for Theoretical Physics,
University of California, Santa Barbara, CA 93106, USA  \\
$^\text{\tiny 3}$Walter Burke Institute for Theoretical Physics, Caltech, Pasadena, CA 91125, USA
\\
$^\text{\tiny 4}$ICTP South American Institute for Fundamental Research, IFT-UNESP, S\~ao Paulo, SP 01440-070 Brazil
}  
\vspace{4mm}
}
\end{center}



\vspace{2mm}
\begin{abstract}
Using integrability, we construct (to leading order in perturbation theory) the explicit form of twist-three light-ray operators in planar $\mathcal{N}=4$ SYM. This construction allows us to directly compute analytically continued  CFT data at complex spin. We derive analytically the ``magic'' decoupling zeroes previously observed numerically. Using the Baxter equation, we also show that certain Regge trajectories merge together into a single unifying Riemann surface. Perhaps more surprisingly, we find that this unification of Regge trajectories is not unique. If we organize twist-three operators differently into what we call ``cousin trajectories'' we find infinitely more possible continuations. We speculate about which of these remarkable features of twist-three operators might generalize to other operators, other regimes and other theories.
%

\end{abstract}

\setcounter{page}{1}
\renewcommand{\thefootnote}{\arabic{footnote}}
\setcounter{footnote}{0}

\newpage


{
\tableofcontents
}



\newpage

\section{Introduction} 

In this work, we explore two universal features of higher dimensional conformal field theories (CFTs) that are naively in tension: (1) there exist a large number of multi-twist local operators \cite{Fitzpatrick:2012yx, Komargodski:2012ek, Fitzpatrick:2015qma}, and (2) CFT data organizes itself into analytic functions of spin \cite{Caron-Huot:2017vep}. For leading double-twist operators these two facts are easy to reconcile. The number of double-twist operators does not grow with spin, and it is easy to connect them together into analytic trajectories. The situation is more complicated for higher-twist trajectories where the number of local operators grows with spin. Before discussing this more complicated case, let us revisit some properties of leading trajectories.
 
 In weakly-coupled CFTs, analyticity in spin of double-twist families can be made fully explicit at the level of operators. Consider a free field $\phi$. The non-local light-ray operator
 \beq
\mathbb{O}_2(S) =  \int_{-\infty}^\infty \dif \alpha_1 \dif \alpha_2 \left(\Psi_{2,S}(\alpha_1, \alpha_2) \equiv {\frac{{|\alpha_1 - \alpha_2|}}{\Gamma(-S)}}^{-1-S}\right) \normord{\phi(\alpha_1 n^+) \phi(\alpha_2n^+)} \label{doubletwistlightray}
 \eeq
(where $n^+$ is a null vector in the $+$ direction)
 is manifestly analytic in $S$. Moreover, for even integer spin $S=2k$, it reduces to the null-integral of the local double-twist spin $2k$ operator $[\phi \phi]_{2k,0}$ defined by
 \beq
 [\phi \phi]_{2k,0} = \normord{\phi \partial_+^{2k}  \phi} +\hspace{3pt} {\partial_+\left( \cdots \right) }\nonumber ,
 \eeq
 with the total derivatives $(\cdots)$ fixed so that $[\phi \phi]_{2k,0}$ is a primary. 
To see this, note that as a distribution,
$\Psi_{2,S}(\alpha_1, \alpha_2) = \delta^{(2k)}(\alpha_1 - \alpha_2) + O(S-2k) \nonumber$. The matrix elements of $\mathbb{O}_2(S)$ therefore provide an analytic continuation in $S$ of the matrix elements of null integrals of $[\phi \phi]_{S,0}$.

Double-twist trajectories are simple because the wavefunction $\Psi_{2,S}$ is completely fixed by kinematics. Requiring that $\mathbb{O}_2$ transforms as a primary operator with dimension $\Delta_L = 1 -S$ translates into demanding that the kernel of (\ref{doubletwistlightray}) is a translation-invariant homogeneous distribution of degree $-1-S$, as we review in section \ref{lightrayconstruction}. $\Psi_{2,S}$ is the unique\footnote{For double-twist trajectories made out of distinct operators $[\phi_1 \phi_2]_{S,0}$ one could consider both ${|\alpha_1-\alpha_2|}^{-1-S}$ and ${\text{sign}(\alpha_1-\alpha_2)|\alpha_1-\alpha_2|}^{-1-S}$. These would correspond to even and odd spin trajectories respectively, see discussion around (\ref{CRTlightray}).} distribution with these properties, up to normalization. For generic $S$, interactions are expected to simply multiplicatively renormalize the operator, leading to anomalous dimensions, see \cite{Caron-Huot:2022eqs} for explicit computations in the Wilson-Fisher theory.\footnote{At special values of $S$, this trajectory can mix with other trajectories, as in the case of the familiar DGLAP-BFKL mixing \cite{Balitsky:2013npa, Brower:2006ea, Jaroszewicz:1982gr, Lipatov:1996ts,Kotikov:2000pm,Kotikov:2002ab}.}


Besides manifesting analyticity in spin, the explicit perturbative expression for the double twist light-ray operators (\ref{doubletwistlightray}) makes it clear that these operators exist in their own right, irrespective of any particular correlator they might contribute to. This is to be contrasted with the known non-perturbative constructions of Regge trajectories  and light-ray operators, which require choices of local operators to build the light-ray operators \cite{Caron-Huot:2017vep,Kravchuk:2018htv}.

It is less clear how to reconcile subleading higher-twist trajectories with analyticity in spin. The number of higher-twist operators grows with spin, with infinitely many operators accumulating in twist at large spin. How should we organize these degenerating operators into analytic families? Do they indeed form discrete trajectories? If so, are these trajectories unique, or do different ways of continuing operators reflect different kinds of Regge limits? How do these infinitely many degenerate trajectories interact with each other at finite spin?

Weakly-coupled theories are an important playground to explore these questions. For example, consider an ansatz for a ``triple-twist'' light-ray operator at weak coupling:
\beq
\mathbb{O}_3(S) =  \int_{-\infty}^\infty \dif \alpha_1 \dif \alpha_2 \dif \alpha_3 \Psi_{3,S}(\alpha_1, \alpha_2, \alpha_3)  \normord{\phi(\alpha_1 n^+) \phi(\alpha_2 n^+) \phi(\alpha_3 n^+) }. \label{tripletwistlightray}
\eeq
The operator $\mathbb{O}_3(S)$ is a primary with dimension $\Delta_L = 1 - S$ for any kernel  $\Psi_{3,S}$ which is translation invariant and homogeneous of degree $-2 - S$. These leads to a continuous number of candidate trajectories analytic in spin, since we can always dress any given kernel with an arbitrary function of, say, $x \equiv {(\alpha_1 - \alpha_2)}/{(\alpha_2 - \alpha_3)}$. What, if anything, quantizes these continuum spectrum of operators down to the discrete Regge trajectories? Is this quantization condition an intrinsically dynamical question whose answer depends very much on the theory under consideration, or is there a more universal property guiding this discretization?

In \cite{Homrich:2022mmd}, we explored the spectrum of multi-twist operators in planar $\mathcal{N}=4$ numerically, finding evidence that  multi-twist families form an infinite number of discretely-spaced Regge trajectories. At any fixed integer spin, only a finite number of these trajectories have nonzero matrix elements, which explains why there are only a finite number of local multi-twist operators at each spin. More recently, \cite{Henriksson:2023cnh} explored this question in the Wilson-Fisher theory, explicitly constructing multi-twist operators and finding discrete Regge trajectories emerge from diagonalizing an appropriate Hamiltonian.

Our goal in this work is to explore the explicit construction of higher-twist single-trace\footnote{Of course, single-trace operators are not quite the universal higher-twist trajectories of general CFTs. However, all of the questions above still apply in the large $N$ context, and many of the lessons we are going to extract are expected to extend to general CFTs.} light-ray operators in planar $\cN=4$ SYM at weak-coupling. %
%
 Key behind the construction is the integrability of the planar dynamics in this very special theory. Requiring that light-ray operators diagonalize higher-integrable charges away from integer spin provides extra constrains on the light-ray kernels. These extra constraints allow us to identify proper quantization conditions which we later reinterpret as more universal requirements on the kernels of light-ray operators. Our results and organization can be summarized as follows:

\begin{itemize}
    \item The kernels are quantized by a smoothness condition when ``partons'' collide, $\alpha_i = \alpha_j$. This is implemented by requiring sufficiently rapid decay of the kernels in Fourier space, equation (\ref{fourierdecay}). Once this condition is imposed, there is a discrete set of operators at each spin diagonalizing the dilatation operator. In planar $\mathcal{N}=4$, to diagonalize the dilatation is equivalent to diagonalizing higher-integrable charges. The latter act as a simple recursion relation in Fourier space, see equation (\ref{recursion})! The quantization condition is then trivial to implement. This is the topic of section \ref{lightrayconstruction}.
    
    \item In section \ref{normsection} we define a \textit{norm} for our light-ray operators, which was first introduced in \cite{Caron-Huot:2013fea}.\footnote{The idea of using this norm was suggested to us by Petr Kravchuk, based on his work on the Wilson Fisher theory \cite{Henriksson:2023cnh}.}  The norm, written in equation (\ref{twopointintegral}), allows us to express the analytic continuation in spin of CFT data in terms of integrals of light-ray kernels, see (\ref{finalformulastructure}). Plugging in our explicit expressions for the perturbative light-ray kernels, we reproduce the continuation of the CFT data for twist-three trajectories obtained in \cite{Homrich:2022mmd}, and derive the ``decoupling zeroes'' uncovered in that reference. The mechanism giving rise to the decoupling zeros is exactly as described in \cite{Henriksson:2023cnh}. 
    
    \item We explore the continuation of twist-three trajectories in the full complex spin plane in section \ref{riemannsurfacesec}. What we observe is rather surprising: the infinitely many trajectories (which are degenerate at tree level), form a rich, infinitely-sheeted Riemann \textit{in perturbation theory}. This is to be contrasted with the traditional mixing of the ``$45$-degree'' trajectories considered here with the horizontal/BFKL and/or shadow trajectories, which  manifests in perturbative CFT data as poles in anomalous dimensions of ever-increasing order \cite{Costa:2012cb,Alfimov:2014bwa,Gromov:2015wca,Klabbers:2023zdz,Caron-Huot:2022eqs}. For higher-twist trajectories, we have infinitely many branch-points at finite positions in the left-half complex spin plane in perturbation theory. For example, we are able to perform the monodromy described in figure \ref{figrmonodromy}: starting from a point corresponding to (the null integral of) a local operator in the second even trajectory\footnote{It turns out that Regge trajectories are ordered in terms of their anomalous dimension for any positive spin.}, we move towards the left half plane, encircle the branch point, and return to the physical region of positive integer spin, now landing on the third even trajectory! Note that, due to the light-ray construction, we can perform this continuation at the level of operators, instead of simply at the level of the spectrum. At any point along the trajectory, one has a well defined operator which might be inserted in arbitrary correlation functions.
    
\item In section~\ref{sec_appendix_newtonconvergence}, we find numerous alternative ways of analytically continuing twist-three operators in spin, which we dub ``CFT cousin trajectories.'' The resulting light-ray operators can be interpreted as twist operators on the string worldsheet. Whether these trajectories persist after nonplanar corrections, and their physical implications, remain open questions.  
    
\end{itemize}

We conclude with a number of open problems and discussion of the various questions raised in this introduction in section \ref{discussionsec}. 

\section{Light-rays and integrability}\label{lightrayconstruction}
\subsection{Review: the light transform}
\label{transformreview}
Light-ray operators are non-local primary operators in a CFT transforming in continuous spin representations \cite{Kravchuk:2018htv} of the conformal group. The simplest light-ray operators can be constructed out of local primary operators via the \textit{light transform}
\beq
\mathbf{L}[\mathcal{O}](x,z)= \int_{-\infty}^{\infty} \dif\alpha (-\alpha)^{-\Delta-S} \mathcal{O}\left(x-\frac{z}{\alpha},z\right). \label{positionlight}
\eeq

Here we use index-free notation, $\mathcal{O}(x,z) \equiv z_{\mu_1} {\makebox[1em][c]{.\hfil.\hfil.}} z_{\mu_S} \mathcal{O}^{\mu_1\dots \mu_S}$ with a future pointing null polarization vector $z$ contracted with the indices of a traceless symmetric tensor operator $\mathcal{O}$. 
The integral moves $\mathcal{O}$ from $x$ towards $\mathscr{I}^+$ along $z$, reaching it at $\alpha = 0$, and then continues into the next Poincare patch in the Lorentzian cylinder, reaching $\mathcal{T}x$ (defined as the point where future-directed light-rays emitted from $x$ converge)  when $\alpha=+\infty$.

The conformal properties of the light-transform are easier to see in the embedding-space \cite{Dirac:1936fq, Mack:1969rr,Ferrara:1972kab,Ferrara:1973yt,Weinberg:2010fx,Costa:2011mg},
\beq
\mathbf{L}[\mathcal{O}](X,Z)=  \int_{-\infty}^{\infty} \dif\alpha (-\alpha)^{-\Delta-S} \mathcal{O}\left(X-\frac{Z}{\alpha},Z\right) =  \int_{-\infty}^{\infty} \dif\alpha  \mathcal{O}\left({Z}-{\alpha}X,-X\right). \label{embeddinglight}
\eeq
Here, we used that $\mathcal{O}$ is a primary operator with dimension $\Delta$ and spin $S$, which in the embedding formalism means $\mathcal{O}(\lambda_1 X,\lambda_2 Z + \beta X) = 
\lambda_1^{-\Delta} \lambda_2^S \mathcal{O}(X,Z)$. In turn, this shows that the light transform $\mathbf{L}[\mathcal{O}]$ transforms as a primary with dimension $\Delta_L = 1 - S$ and spin $J_L = 1-\Delta$ since
\beq
\mathbf{L}[\mathcal{O}](\lambda_1 X, \lambda_2 Z) = \lambda_1^{-1+S}\lambda_2^{1-\Delta} \mathbf{L}[\mathcal{O}](X,Z). \nonumber
\eeq
The embedding space formula (\ref{embeddinglight}) also manifests that $\alpha = 0 $ is a regular point of the integral transform. 

Equation (\ref{positionlight}) can be recovered from (\ref{embeddinglight}) through the standard Poincare gauge fixing $X = (1, x, x^2)$ and $Z=\left(0,2 x\cdot z, z\right)$. For our purposes, it will be more useful to choose coordinates so that the light-ray fits in a single Poincare patch, stretching from $\mathscr{I}^-$ to $\mathscr{I}^+$. This is achieved by setting \beq
X=(0,-2 y \cdot z, -z) \qquad \text{and} \qquad Z = (1, y, y^2), \label{oursectioncoordinates}\eeq which results in
\beq
\mathbf{L}[\mathcal{O}](x',z) = \int \dif\alpha \mathcal{O}(y + \alpha z, z), \nonumber
\eeq
with $x' = y - \infty z$.

The light transform of even and odd spin operators can be distinguished through $CRT$ symmetry, which acts on local operators as
$U_{\textbf{CRT}}\mathcal{O}(x,z) U_{\textbf{CRT}}^{-1} = (-1)^S \mathcal{O}^\dagger(\bar{x},z),$
where  $\bar{x} = (-x_0, -x_1, x_2, \dots x_d)$. This results in
\beq
\left(U_{\textbf{CRT}} \mathbf{L}[\mathcal{O}](x',z) U_{\textbf{CRT}}^{-1}\right)^\dagger = (-1)^S \mathbf{L}[\mathcal{O}](x',z). \label{CRTlighttransform}
\eeq
When constructing continuations in spin of light transforms, we will require that even/odd spin trajectories at generic spin $S$ are eigenvalues of CRT plus hermitian conjugation with eigenvalues $\pm 1$ respectively.

\subsection{Light-ray operators at weak coupling}
\label{weakcouplingsec}
At weak coupling, one might propose to construct more general light-ray operators in terms of fundamental fields. For example, consider the following ansatz in $\mathcal{N}=4$ SYM:
\beq
\mathbb{O}(X,Z) = \int_{-\infty}^{\infty} \prod_{i=1}^L \dif \alpha_i \Psi(\alpha_1,\dots, \alpha_L) \text{Tr}\left(\mathcal{Z}(Z -\alpha_1 X)\dots \mathcal{Z}(Z -\alpha_L X) \right). \label{ansatzhighertwist}
\eeq
We use the trace to denote Wilson lines along the contour cyclically connecting the various field insertions. Due to the cyclicity of the trace, it is enough to consider cyclic-invariant kernels. For $\Psi$ given by appropriate linear combinations of (derivatives of) Dirac delta functions, equation (\ref{ansatzhighertwist}) reduces to the light-transform of local single-trace operators in the $\text{SL}\left(2,\mathbb{R}\right)$ sector of $\mathcal{N}=4$ SYM, schematically given by
\beq
\text{Tr}\left(D_+^S\mathcal{Z}^L\right) + \texttt{permutations} \label{localsingletraces}.
\eeq
The ansatz (\ref{ansatzhighertwist}) produces a tree-level primary operator with dimension $\Delta_L = 1 - S$ and spin $J_L = 1 - L - S$, provided that $\Psi$ is a translation-invariant distribution with homogeneity degree $J_L$. We will often refer to $\Psi$ as the light-ray wave function. For fixed $(\Delta_L,J_L)$, there is a continuum of such wave functions. As discussed in the introduction, our first goal is to understand how (and if) the dynamics of the theory selects a preferred discrete set of light-ray operators (\ref{ansatzhighertwist}) corresponding to the Regge trajectories of the multi-twist operators  (\ref{localsingletraces}). 

Henceforth, we use the coordinates defined in equation (\ref{oursectioncoordinates}). Even- and odd-spin trajectories will correspond to independent analytic operators. Following equation (\ref{CRTlighttransform}), it is natural to consider operators with definite signature under
CRT plus hermitian conjugation, i.e.\ such that
\beq
\left(U_{\textbf{CRT}} \mathbb{O}(x',z) U_{\textbf{CRT}}^{-1}\right)^\dagger = \pm \mathbb{O}(x',z), \label{CRTlightray}
\eeq
with the upper/lower sign corresponding to even/odd spin trajectories. In terms of the wave functions, the requirement (\ref{CRTlightray}) corresponds to $\Psi(\alpha_1,\dots, \alpha_L) = \pm \Psi(-\alpha_1,\dots, -\alpha_L)$. 

Let us summarize the kinematic constrains on the light-ray wave functions.
\begin{center}
\begin{minipage}{0.9\textwidth}
\begin{mdframed}[style=MyFrame]
\begin{center}
    \textbf{{Kinematic Constrains on $\Psi$}}
    \end{center}
    \vspace{0.3cm}
    The wave function $\Psi$ of single-trace light-ray operators
    \beq
\mathbb{O}(x',z) = \int_{-\infty}^{\infty} \prod_{i=1}^L \dif \alpha_i \Psi(\alpha_1,\dots, \alpha_L) \text{Tr}\left(\mathcal{Z}(y +\alpha_1 z)\dots \mathcal{Z}(y + \alpha_L z) \right) \label{ansatzmultitwist}
    \eeq
    continuing the light transform of single-trace $\text{SL}\left(2,\mathbb{R}\right)$ operators (\ref{localsingletraces}) must be
\begin{align}
&\bullet\label{tinv} \text{ Translation invariant: }&\Psi(\alpha_1, \dots, \alpha_L) &= \Psi(\alpha_1 + a, \dots, \alpha_L + a),\\ &\bullet\text{ Homogeneous: }&\Psi(\lambda \alpha_1, \dots, \lambda \alpha_L) &= \lambda^{1-L-S}\Psi(\alpha_1, \dots, \alpha_L), \label{hinv}\\
&\bullet\text{ Reflection symmetric: }&\Psi(-\alpha_1, \dots,  -\alpha_L) &= \pm \Psi(\alpha_1, \dots, \alpha_L) \label{rinv},
\\&\bullet\text{ Cyclic symmetric: }&\Psi(\alpha_1, \dots,  \alpha_L) &= \Psi(\alpha_2, \dots, \alpha_1)\label{cinv}.
\end{align}
\end{mdframed}
\end{minipage}
\end{center}

It is important to trivialize the kinematic conditions (\ref{tinv}-\ref{cinv}) as much as possible before considering the dynamics. We will first consider in detail the trivial case $L=2$ and then the $L=3$ case, which is the main case of interest in this work. 

As described in the introduction, the twist-2 wave function is completely fixed by the kinematic constrains (\ref{tinv}-\ref{cinv}). Indeed, translation invariance implies that, in this case,
\beq
\nonumber \Psi(\alpha_1, \alpha_2) = f(\alpha_1 - \alpha_2).
\eeq
Homogeneity now requires that $f$ is an homogeneous distribution of degree $-1-S$. These are classified to be of the form
\beq
f(\alpha_1-\alpha_2) = c_1(\alpha_1 - \alpha_2 + i \epsilon)^{-1-S}+c_2(\alpha_2 - \alpha_1 + i \epsilon)^{-1-S}. \nonumber
\eeq
For even spin trajectories\footnote{There are no odd spin twist-2 operators in the  $\text{SL}\left(2,\mathbb{R}\right)$  sector.}, reflection symmetry then fixes $c_1 = c_2$. Let us normalize $c_1 = 1/\Gamma(-S)$ so as to cancel the singularities of $\Psi$ at positive even integer values of $S$. With this normalization, $\Psi$ is holomorphic\footnote{This means that, for any test function $h$, $\langle \Psi, h \rangle $ is a holomorphic function of $S$.} for all $S\in \mathbb C$. In the end, we have
\beq
\nonumber \Psi(\alpha_1, \alpha_2) = {\frac{|\alpha_1 - \alpha_2|}{\Gamma(-S)}}^{-1-S},
\eeq
which is to be compared with equation (\ref{doubletwistlightray}).

Let us now move on to the case $L=3$. It is useful to introduce the polar coordinates $(r, \theta)$ defined through
\begin{align}
\alpha_1 - \alpha_2 &= r \cos(\theta), \label{alphachange1}\\
\alpha_2 - \alpha_3 &= r \cos(\theta-2\pi/3),\\
\alpha_3 - \alpha_1 &= r \cos(\theta-4\pi/3)\label{alphachange3},
\end{align}
the last equation being automatic. The primary condition (\ref{tinv}) implies that the wave function $\Psi$ is a function of $r$ and $\theta$ only, being independent of the center of mass coordinate $3 \alpha_\text{cm} \equiv \alpha_1 + \alpha_2 + \alpha_3$. In terms of these coordinates, homogeneity (\ref{hinv}) takes the simple form
\beq
\Psi(\alpha_1, \alpha_2, \alpha_3) = \frac{g(\theta)}{\Gamma(-S) r^{2+S}}. \label{rthetaansatz}
\eeq
where $g(\theta)$ is an undetermined distribution on $S^1$ and the $\Gamma(-S)$ factor is added for later convenience. The discrete symmetries (\ref{rinv}-\ref{cinv}) then reduce to 
\beq
    g(\theta)=g(\theta+ \pi), \qquad g(\theta) = g(\theta + 2\pi/3), \nonumber
\eeq
which can be combined into
\beq
g(\theta) = g(\theta + \pi/3). \label{discretesym}
\eeq

\subsection{Dynamics}

The undetermined wave functions $g(\theta)$ are intrinsically dynamical objects. They should be constrained     by requiring that the light-ray operators are eigenvectors of the dilatation operator $\mathbb{D}$ in the interacting theory, much like how $\mathbb{D}$ determines the precise permutations in (\ref{localsingletraces}) for the various single-trace primary operators at the interacting level.

 In planar $\mathcal{N}=4$ SYM, the task of diagonalizing the action of the dilatation operator is greatly simplified due to integrability. In this theory, primary single trace operators are not only eigenstates of the dilatation operator, but also carry higher quantum numbers corresponding to extra integrable charges. This is a well known story at the level of local operators. Here we simply demand this remains true at the level of single-trace light-ray operators.

At one-loop order, these higher-charge operators can be constructed explicitly and take the form of simple differential operators acting on the various field insertions along the Wilson line \cite{Faddeev:1996iy, Beisert:2004ry}. Their common building block is the R-matrix operator, whose action on the  $\mathcal{Z}$ fields of (\ref{ansatzmultitwist}) is
\beq
R_{a b}\cdot \mathcal{Z}(\alpha_j ) 
= \begin{bmatrix}
u + \tfrac{i}{2}\left(\alpha_j \partial_{\alpha_j} + \tfrac{1}{2} \right) & - \tfrac{i}{2} \partial_{\alpha_j} \vspace{0.1cm}\\
\tfrac{i}{2}\left(\alpha_j^2\partial_{\alpha_j} + \alpha_j\right) & u - \tfrac{i}{2}\left(\alpha_j \partial_{\alpha_j} + \tfrac{1}{2} \right)
\end{bmatrix}_{ab} \cdot \mathcal{Z}(\alpha_j),
\eeq
where we denote $ \mathcal{Z}(\alpha_j) \equiv \mathcal{Z}(y + \alpha_j z)$ and set $z=n^+$ for simplicity.

\begin{figure}[t]
\centering
\includegraphics[width=\textwidth]{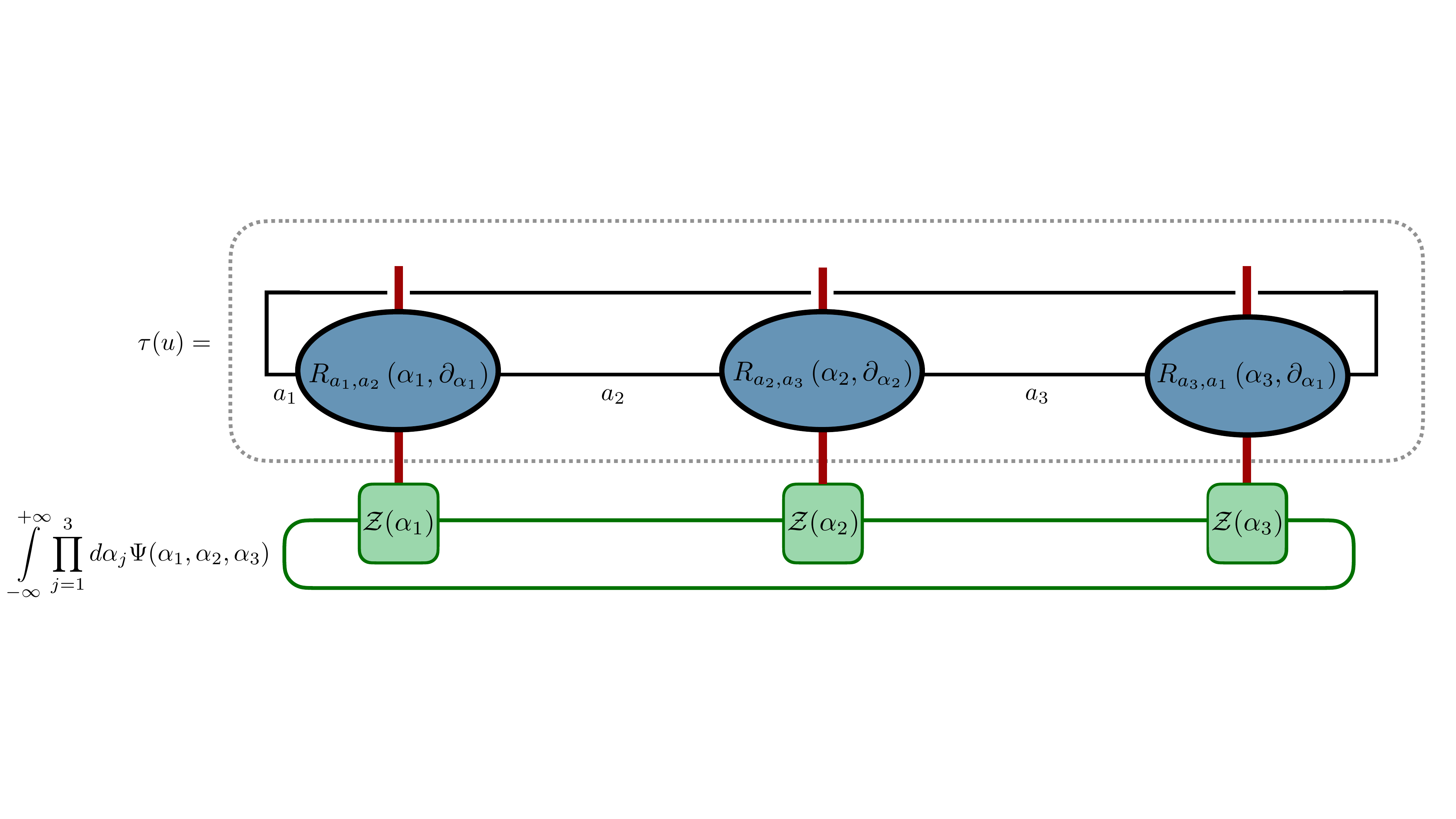}
\vspace{-3cm}
\caption{Action (\ref{taudef}) for $L=3$. In this case, each $R$-matrix is linear in $u$ so in total $\tau(u)$ is a differential operator in $\alpha_j, \partial_{\alpha_j}$ which is cubic in $u$; in other words, it is a collection of four differential operators, one for each power of $u$. In practice three of these are trivial and one leads to a nice differential equation on the wave function upon integration by parts. This leads to equation (\ref{difeqL3}) coming from the $u^0$ term in $\tau(u)$.}
\label{figtransfer}
\end{figure}

From the R-matrix, one can construct the transfer-matrix, which acts on the full light-ray operator through
\beq
\tau(u)\cdot \text{Tr}\left(\mathcal{Z}(\alpha_1 z)\dots \mathcal{Z}( \alpha_L) \right) = \text{Tr}\left(\left(R_{a_1 a_2} Z(\alpha_1) \right)\dots \left(R_{a_L a_1} Z(\alpha_L) \right)\right), \label{taudef}
\eeq
see figure \ref{figtransfer} for a graphical description of $R$ and $\tau$. For our purposes, it suffices to know that the transfer matrix commutes with itself and the dilatation operator for any value of the spectral parameter,
\beq
[\tau(u),\tau(v)] = [\tau(u), \mathbb{D}] = 0.
\eeq

Local single-trace operators are eigenvectors of $\tau(u)$ and therefore so are their light transforms. We demand that this remains true for more general light-ray operators corresponding to the continuation of the light-transforms at non-integer spin:
\beq
\tau(u)\cdot \mathbb{O} = t(u) \mathbb{O}. \label{transfermatrixequation}
\eeq

The transfer matrix equation (\ref{transfermatrixequation}) is a polynomial in $u$ of degree $L$. Collecting (\ref{transfermatrixequation}) in powers of $u$ provides a number of differential equations that must be satisfied by $\mathbb{O}$. The differential operators naturally act on the partons through (\ref{taudef}). Integrating the action  by parts results in a system of differential equations for the wave functions $\Psi$. Ignoring the trivial $u^L$ term, this provides a set of $L$ equations on the $L$-variable wavefunction $\Psi$. Note that, after imposing the kinematical constrains, the $u^{L-1}$ and $u^{L-2}$ terms in (\ref{transfermatrixequation}) are redundant, the first being equivalent to (\ref{tinv}) and the second\footnote{The $u^{L-2}$ term in $\tau(u)$ is proportional to the quadratic Casimir of the the SL$(2,\mathbb{R})$ group of conformal symmetries that preserve the light-ray. } equivalent to (\ref{hinv}). We are thus left with $L-2$ non-trivial conditions on $\Psi$. For $L=3$, using (\ref{rthetaansatz}), the nontrivial equation reads 
\begin{align}
&\tfrac{1}{2}\left(1 + \cos(6 \theta)\right) g^{(3)} - \tfrac{3}{2}(4+S) \sin(6 \theta)g^{(2)} -\left( \left(3 S^2 + 6 S + 8 \right) + \left(3S^2 + 24S +44 \right)\cos(6\theta)\right) g'  \nonumber \\
&= (\tfrac{1}{2} (12 i \sqrt{3} q + (2 + S) (4 + S) (6 + S) \sin(6 \theta))) g , \label{difeqL3}
\end{align}
where $q$ is defined by \beq
t(u) = 2u^3 - \tfrac{1}{4}C_2(S) u + \tfrac{q}{8}\label{eigentau},\eeq with $C_2(S) = S^2 + 2S +\tfrac{3}{2}$.

The periodicity of (\ref{difeqL3}) invites us to work in Fourier space. The discrete symmetries of $g$, equation (\ref{discretesym}), imply that, as a distribution\footnotemark, 
\beq
g(\theta) = \sum_{n \in \mathbb{Z}} a_{n} (-1)^n  e^{i 6 n \theta}, \label{fourierdec}
\eeq
where we introduced factors of $(-1)^n$ for later convenience. Indeed, plugging the decomposition (\ref{fourierdec}) in (\ref{difeqL3}) leads to a beautiful three-term recursion relation! We summarize the results of this section in the following box.

\begin{center}
\begin{minipage}{0.9\textwidth}
\begin{mdframed}[style=MyFrame]
\begin{center}
    \textbf{{Dynamic Constraints on $\Psi$ ($L=3$)}}
    \end{center}
    \vspace{0.3cm}
    The $g(\theta)$ distribution defining one-loop even-trajectory single-trace light-ray operators through (\ref{rthetaansatz}) must satisfy
\begin{align}
&  (6n - S)(2+6n - S)(4 +6n-S) a_{n+1}  \nonumber \\ &-12(n\left(8 + 36 n^2 +6 S + 3 S^2 \right) - 2 \sqrt{3} q) a_{n} \label{recursion}\\
&+( 6n -4 + S)(6n -2 + S)(6n+S) a_{n-1} \nonumber =0,  
\end{align}
    where $a_n$ are the Fourier coefficients of $g(\theta)$ defined in (\ref{fourierdec}) and $q$ is
    an operator dependent charge defined in (\ref{eigentau}).
\end{mdframed}
\end{minipage}
\end{center}

The determination of the light-ray wave functions corresponding to the various presumed discrete trajectories can be separated into two parts. First, one should understand what fixes, for the various discrete trajectories, indexed by $j$, the integrable charge eigenvalue $q_j(S)$ at non-integer spin. Second, given $q_j(S)$, one must determine how to fix a unique solution to (\ref{recursion}). At this point
there remains a continuous one parameter family of solutions to (\ref{recursion}). We solve these issues in section \ref{sec_quantization}. However, first it is useful to consider the case of light transforms of local operators from the perspective of (\ref{recursion}).

\footnotetext{This means that (\ref{fourierdec}) might not converge pointwise, but does when integrated against a test function $f$ whose Fourier coefficients $c_n$ decay faster than any power, i.e.\ \beq
\langle g, f \rangle \equiv \int_{0}^\infty \dif\theta g(\theta)f(\theta) = \sum_n a_n c_{-6n} (-1)^n \label{convdist}
\eeq converges. Equivalently, a ($\pi/3$ periodic) distribution on $S_1$ is defined by a sequence of polynomially bounded coefficients $a_n$ through the right hand side of  (\ref{convdist}).}


\subsection{Integer spin and reduction to the light transform}
\label{reductionsection}
 For even integer spin $2k$, the ansatz (\ref{ansatzmultitwist}, \ref{rthetaansatz}) reduces to
\beq
\mathbb{O} = \int \dif \alpha_{cm} \dif \theta \dif r g(\theta) \delta^{(2k)}(r) \text{Tr}\left(\mathcal{Z}(y +\alpha_1 z)\mathcal{Z}(y +\alpha_2 z)\mathcal{Z}(y +\alpha_3 z)\right) + O(S-2k), \nonumber 
\eeq
where we assumed that $g(\theta)$ is regular as $S=2k$, and the $\alpha_i$ are understood to be functions of $\alpha_{cm}, r, \theta$ according to (\ref{alphachange1}-\ref{alphachange3}). Because the radial part of the kernel degenerates into a delta function the fields $\mathcal{Z}$ localize,
\beq
\mathbb{O} = \int \dif \alpha_{cm} \dif \theta  g(\theta)\sum_{j=0}^{2k} P_j(\theta) \text{Tr}\left(\mathcal{Z}(z\cdot D)^j\mathcal{Z}(z\cdot D)^{2k-j}\mathcal{Z}\right)(y +\alpha_{cm} z) + O(S-2k), \label{integerloc}    
\eeq
where $P_j$ is a homogeneous polynomial of degree $S$ on sine and cosine of $\theta$. Due to this property of $P_j$, the wave function $g(\theta)$ can be truncated  in its Fourier decomposition:
only $a_n$ with $|n| \leq \lfloor\tfrac{S}{6} \rfloor$ can contribute in (\ref{integerloc}).

Let us now examine the recursion relation (\ref{recursion}) for integer spin. Beautifully, for even $S$ the equations indexed by $|n|\leq \lfloor\tfrac{S}{6} \rfloor$ close! Indeed, these equations provide a system of $2\lfloor\tfrac{S}{6}\rfloor + 1$ equations on the $2\lfloor\tfrac{S}{6}\rfloor + 1$ variables $a_n$, $|n| \leq \lfloor\tfrac{S}{6} \rfloor$. A non-trivial solution to these linear equations is possible if and only if the determinant of the matrix of coefficients vanishes. This condition is a polynomial equation of degree $2\lfloor\tfrac{S}{6}\rfloor + 1$ for $q$. Each solution to this equation will then define, up to normalization, the light transform of a local operator by means of (\ref{integerloc})! Indeed, there are $2\lfloor\tfrac{S}{6}\rfloor + 1$ primary twist three operators at even spin $S$. (See appendix \ref{countingoperators} for the counting of local primaries for arbitrary twist $L$ and spin $S$.)  We therefore recover the complete spectrum of even-spin twist-three local operators from the light-ray construction, including the integrable charge $q$ characterizing each of them.\footnote{Bethe ansatz provides an alternative way of determining the spectrum of $q$ at a given integer spin $S$, see e.g.\ section IV of \cite{Homrich:2022mmd}. }

\begin{figure}[t!]
\centering
\includegraphics[width=\textwidth]{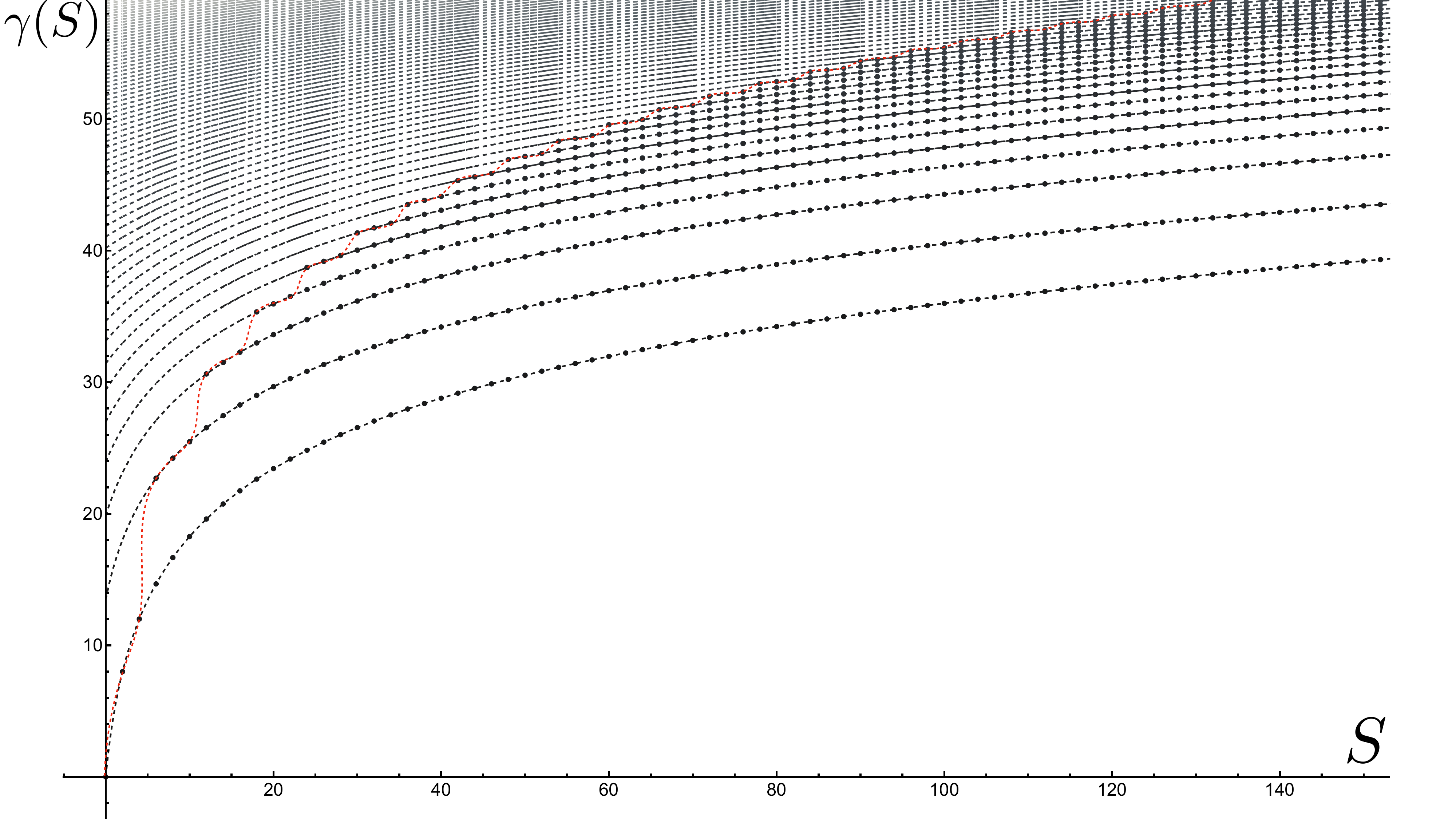}
\vspace{-0.3cm}
\caption{Black dots: even spin twist three spectrum up to $S=150$. Black dashed curves: first $50$ analytic trajectories continuing the even spin spectrum. They correspond to the light-ray operators constructed in section \ref{lightrayconstruction}. The black lines are constructed entirely from the the physical states using (\ref{newtonseries}). Alternatively, they can be obtained from the action of the light-ray Hamiltonian, see discussion around (\ref{tildegamma}), or directly from integrability, see section \ref{riemannsurfacesec}. One might wonder if there are alternative ways of organizing the spectrum into analytic trajectories. Consider for example the red dashed curve, which is drawn by hand. Is there an analytic trajectory which continues the selected local operators? We expect that the answer is negative, and indeed, under the assumption of analyticity in a half-plane, this can be verified using (\ref{newtonseries}), see appendix \ref{sec_appendix_newtonconvergence}. However, surprisingly, it turns out that, indeed, there are (infinitely many) other reorganizations of the spectrum into analytic families. We discuss this surprising fact in the conclusions and in appendix \ref{sec_appendix_newtonconvergence}.}
\label{CFTdatafig}
\end{figure}

Due to integrability, it is straightforward to compute the CFT data associated to these operators. In figure \ref{CFTdatafig} we present all\footnote{Except for the lowest operator at each spin, operators come in degerante pairs, hence why at even spin $S$ there are $\lfloor\tfrac{S}{6}\rfloor +1$ points in figure \ref{CFTdatafig}.} one-loop anomalous dimension for even spin twist three operators up to spin $S=150$. 

The black dashed lines in figure \ref{CFTdatafig} represent a complete set of trajectories ordered by their anomalous dimensions.  It is possible to provide a numerical argument that this assignment of local operators into families indeed corresponds to \textit{analytic} Regge trajectories.
Assuming no mixing of trajectories in a half right plane\footnote{In the case of the black trajectories in figure \ref{CFTdatafig}, it is verified a posteriori, see \cite{Homrich:2022mmd}, that there are indeed no branch points for  $\RE(S)\geq0$.}, the complex spin CFT data associated to a given Regge trajectory admits a convergent expansion in terms of the integer spin CFT data through a Newton series \cite{Homrich:2022mmd}. For instance, in the case of the anomalous dimension $\gamma(S)$ we have
\beq
\gamma(S) = \sum_{n=0}^\infty \binom{\tfrac{S-S_0}{2}}{n} \sum_{i=0}^n (-1)^{n-i} \binom{n}{i}\gamma_{S_0 + 2 i}, \label{newtonseries}
\eeq
where $\gamma_S$ corresponds to the integer spin CFT data. Convergence is guaranteed to the right of the first singularity in the complex spin plane. In (\ref{newtonseries}), we are free to choose $S_0$ to be any integer along the trajectory --- $\gamma(S)$ should be insensitive to this choice for a consistent Regge trajectory. One can then use the self-consistency of equation (\ref{newtonseries}) to probe whether a given assignment of local operators into trajectories is consistent with analyticity in spin.

Plugging the anomalous dimension associated to the integer spin operators of the black trajectories in figure \ref{CFTdatafig} indeed results in beautiful numerical convergence of (\ref{newtonseries}). Indeed, the black curves themselves are generated through Newton's formula,\footnote{We use CFT data up to spin $S=500$ in Newton's formula, which allows us to control the higher trajectories whose local operators are not included in figure \ref{CFTdatafig}.} and can be double checked against other methods, see \cite{Homrich:2022mmd}. On the other hand, alternative assignments of local operators to trajectories, such as the red curves in figure \ref{CFTdatafig}, result in numerical divergences via (\ref{newtonseries}) --- see section \ref{sec_appendix_newtonconvergence}. This suggests that such assignments do not correspond to Regge trajectories controlling CFT correlators.

Newton's series can also be used to reconstruct the analytic continuation of the integrable charge $q_j(S)$ associated to the various black trajectories, indexed henceforth by $j$, from the integer spin $q$ values obtained from (\ref{recursion}). The problem of determining the values of $q_j(S)$ for non-integer spin is thus in practice solved, at least in a half-plane. In this work, however, we would like to understand how the black trajectories and $q_j(S)$ are selected from the point of view of light-ray operators, i.e.\ directly from (\ref{recursion}) at non-integer spin. 

\subsection{Quantization at non-integer spin}
\label{sec_quantization}

Light-ray operators at non-integer spin should diagonalize the one-loop dilatation operator, which acts by
\begin{align}
  &  \mathbb{D}_1  \cdot \text{Tr}\left(\mathcal{Z}(\alpha_1)\dots \mathcal{Z}(\alpha_L ) \right) = \sum_{i=1}^L \mathcal{H}_{i,{i+1}}\cdot \text{Tr}\left(\mathcal{Z}(\alpha_1 )\dots \mathcal{Z}(\alpha_L ) \right) \label{dilatationdef},
  \\ & \nonumber
    \mathcal{H}_{i,{i+1}}\cdot \text{Tr}\left(\mathcal{Z}(\alpha_1 )\dots \mathcal{Z}(\alpha_L ) \right) = 2g^2\int \frac{\dif\tau}{\tau} \Big(2 \text{Tr}\left(\dots \mathcal{Z}(\alpha_i ) \mathcal{Z}(\alpha_{i+1} ) \dots \right)  
    \\&  - \text{Tr}\left(\dots \mathcal{Z}((1-\tau) \alpha_i + \tau \alpha_{i+1}) \mathcal{Z}(\alpha_{i+1} ) \dots \right)- \text{Tr}\left(\dots  \mathcal{Z}(\alpha_{i} )\mathcal{Z}((1-\tau) \alpha_{i+1} + \tau \alpha_{i}) \dots \right) \Big)\nonumber,
\end{align}
where $\mathcal{Z}(\alpha) \equiv \mathcal{Z}(\alpha n^+)$.
Demanding that $\mathbb{O}$ diagonalize $\mathbb{D}_1$ implies an integral equation for the light-ray wavefunction $\Psi$. In the study of the Wilson-Fisher theory in \cite{Henriksson:2023cnh}, an analogous integral equation implied certain smoothness/boundary conditions for $\Psi$, leading to quantization conditions and discrete solutions.

Here, instead of analyzing how smoothness/boundary conditions come about from (\ref{dilatationdef}), we will instead guess a simple quantization condition using our Fourier-space recursion relation, and then check its consistency with $\mathbb{D}_1$. It is natural to expect that the quantization condition is associated to analyticity properties of $g(\theta)$. We are thus led to examine the asymptotics of the Fourier coefficients $a_n$.

Let us consider the large $|n|$ asymptotics of (\ref{recursion}) and look for power series  solutions. The most general asymptotic solution takes the form
\beq
a_n = \alpha_1 n^{-1} \left(1 + \sum_{i=1}^\infty a_i n^{-i}\right)+\alpha_2 n^{S} \left(1 + \sum_{i=1}^\infty b_i n^{-i}\right),  \label{asymptotics}
\eeq
with $a_i$, $b_i$ fixed in terms of $q$ and $S$ through the recursion relation. Let us emphasize that the asymptotic behaviors at $n\rightarrow \pm \infty$ are in principle distinct. Both should take the form~(\ref{asymptotics}), but with different values of $\alpha_i$. Let us denote those by $\alpha_i(\pm \infty)$ respectively. It is a non-trivial connection problem to determine $\alpha_i(-\infty)$ given the $\alpha_i(+\infty)$ asymptotics, $q$ and~$S$.

 This connection problem can be explored numerically. Truncating the sums in (\ref{asymptotics}) with a decent number of terms provides a good approximation to a solution of (\ref{recursion}) at large positive values of $n$. From there, one can reconstruct the solution for arbitrary $n$ through iterations of the recursion relation (\ref{recursion}). In this way, one can numerically fit $\alpha_i(-\infty)$ as a function of $q$, $S$, and $\alpha_i(+\infty)$.

 \begin{figure}[t!]
\centering
\includegraphics[width=1\textwidth]{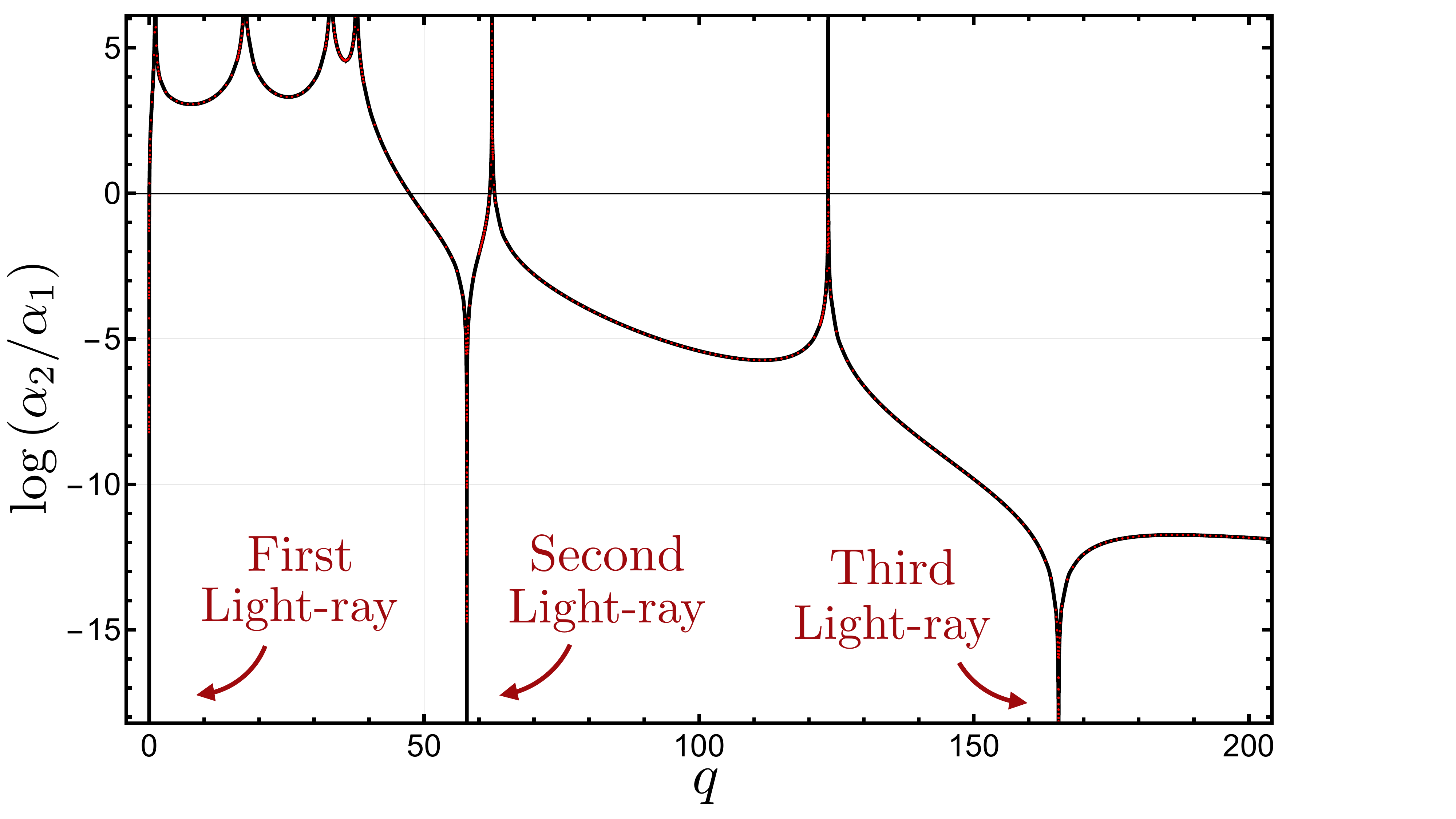}
\vspace{-0.3cm}
\caption{We consider solutions of the recursion relation (\ref{recursion}) that decay as $n^{-1}$ for $n\rightarrow+\infty$. The recursion relation then determines the asymptotics as $n \rightarrow -\infty$ to be of the form (\ref{asymptotics}) with $\alpha_2/\alpha_1$ fixed in terms of $S$ and $q$. Here we plot the ratio as a function of $q$ for $S=6+1/3$. Zeroes of the ratio correspond to values of $q$ for which solutions with pure $|n|^{-1}$ asymptotic exist. Our main proposal is that these are precisely the light-ray operators corresponding to Regge trajectories, see (\ref{fourierdecay}).}
\label{connectionfig}
\end{figure}

 In figure \ref{connectionfig} we plot the ratio ${\alpha_2(-\infty)}/{\alpha_1(-\infty)}$ as a function of $q$ for the solution of (\ref{recursion}) with $\alpha_2(+\infty) = 0$. We fix the spin $S=6+1/3$, just for the purposes of illustration. Zeroes of the ratio correspond to discrete values $q_j(S)$ for which solutions with pure $|n|^{-1}$ asymptotics --- and thus maximal smoothness in $\theta$ space --- exist. The result is remarkable: these discrete values correspond precisely to the values $q_j(S)$ determined by the black trajectories in figure \ref{CFTdatafig}! We are thus led to the following proposal.

\begin{center}
\begin{minipage}{0.9\textwidth}
\begin{mdframed}[style=MyFrame]
\begin{center}
    \textbf{{Quantization Condition on $\Psi$ ($L=3$)}}
    \end{center}
    \vspace{0.3cm}
   The solutions of (\ref{recursion}) corresponding to light-ray operators governing Regge trajectories, equivalent to the black curves in figure \ref{CFTdatafig}, are those for which asymptotically
   \beq a_n \sim |n|^{-1}. \label{fourierdecay}
   \eeq
   At fixed $S$, these exist only for discrete values $q_j$ of the integrable charge. These define the infinitely many discrete twist three trajectories as a function of spin.
\end{mdframed}
\end{minipage}
\end{center}

 \begin{figure}[t!]
\centering
\includegraphics[width=0.8\textwidth]{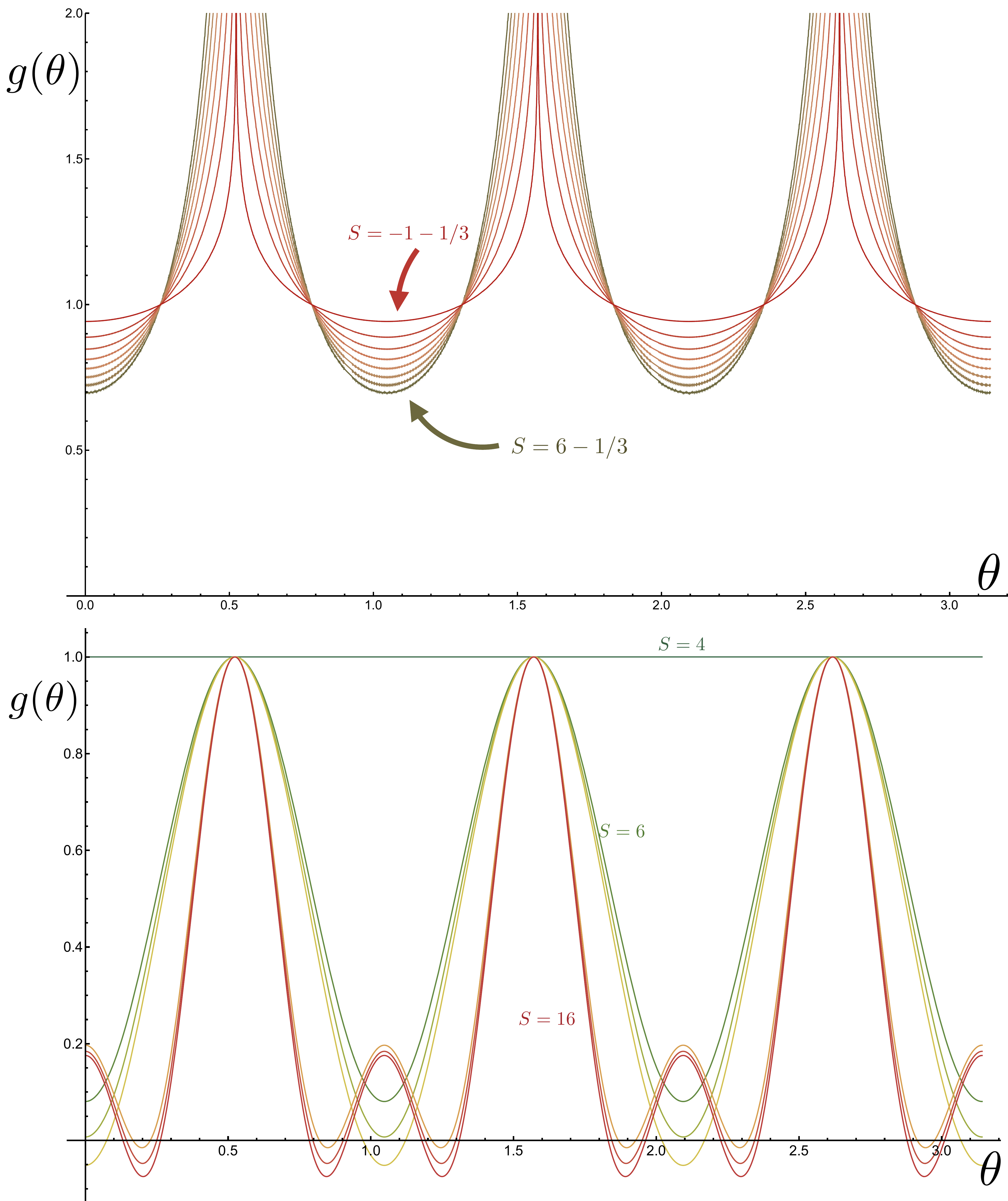}
\caption{Top: The dynamical part of the wave function, $g(\theta)$, for the lowest even-spin trajectory at various values of spin. From red to green, spin is taken from $S=-1-1/3$ to $S=6-1/3$ in unit steps. The wave function has logarithmic divergences when partons collide. We see three such divergencies in the plot, which correspond respectively to~$\alpha_2 = \alpha_3$,~$\alpha_{1}=\alpha_2$,  and $\alpha_3 = \alpha_1$. 
Bottom: At integer spin, it is enough to truncate the wave function to the first few Fourier modes in order to recover the light-transform of local operators, see section \ref{reductionsection}. From green to red, the lowest trajectory at even spin from $S=4$ to $S=16$. 
}
\label{plotofwave functions}
\end{figure}

To test the proposal (\ref{fourierdecay}) we can check the action of the one-loop dilatation operator (\ref{dilatationdef}) on the corresponding light-ray operators. 
In practice, we compute
\beq
\tilde{\gamma} \equiv  \frac{\langle \Omega | \Tr\left(\bar{\mathcal{X}}^2\bar{\mathcal{Z}}\right)(x_1)  \[\mathbb{D}_1, \mathbb{O}(x_2) \]
\Tr\left({\mathcal{X}} \bar{\mathcal{Z}}\right)(x_3) \Tr\left({\mathcal{X}} \bar{\mathcal{Z}}\right)(x_4) |\Omega \rangle}{\langle \Omega | \Tr\left(\bar{\mathcal{X}}^2\bar{\mathcal{Z}}\right)(x_1) \mathbb{O}(x_2)
\Tr\left({\mathcal{X}} \bar{\mathcal{Z}}\right)(x_3)
\Tr\left({\mathcal{X}} \bar{\mathcal{Z}}\right)(x_4) |\Omega \rangle} \label{tildegamma}
\eeq
for a range of cross ratios $z, \bar{z}$. We observe that solutions defined through the criteria (\ref{fourierdecay}) indeed correspond to cross-ratio independent $\tilde{\gamma}$, as it must be for light-rays diagonalizing $\mathbb{D}_1$. Solutions of (\ref{recursion}) with alternative asymptotics do not pass this consistency check, $\tilde{\gamma}$ being a non-trivial function of the cross-ratio in those cases. Moreover, for the solutions defined by (\ref{fourierdecay}), we observe that $\tilde\gamma$ agrees with $\gamma(S)$, as determined by Newton series, for each of the black trajectories\footnote{\label{lowesttfoonote} The lowest trajectory in figure \ref{CFTdatafig} provides a nice more analytic testing ground in which one can disentangle the determination of $q$ and of the asymptotics. This trajectory possess enhanced symmetry. It is invariant under inversion of the orientation of the Wilson line in (\ref{ansatzmultitwist})  - as can be verified explicitly for the various light-transforms/local operators along the trajectory - or, equivalently, the wave function is invariant under permutations:
\beq
\Psi(\alpha_1, \alpha_2, \alpha_3) = \Psi(\alpha_2,\alpha_1,\alpha_3) = \Psi(\alpha_2, \alpha_3,\alpha_1). \label{fullpermsymmetry}
\eeq
This property is equivalent to the condition $q=0$ for all $S$, as can be verified from the $n \leftrightarrow -n$ symmetry of the recursion relation (\ref{recursion}). One does not need to compute $q(S)$ numerically in this case. Solutions to the recursion for $q=0$ automatically have symmetrical asymptotics $\alpha_i(\infty) = \alpha_i(-\infty)$. Using the solution with pure $n^{-1}$ asymptotics in (\ref{tildegamma}) reproduces the correct anomalous dimension for these lowest lying trajectory,
\beq
\gamma_{q=0}(S) = 8 h(S/2).
\eeq
For the higher trajectories the anomalous dimensions are not known in closed form, but we observe a beautiful match when comparing with the continuations determined from Newton's series (\ref{newtonseries}). } in figure \ref{CFTdatafig}.

We have thus completed the explicit construction of the \textit{operators} analytically continuing the (light transform of) even\footnote{The construction for odd spin operator is identical, with the minor modification that in the Fourier decomposition (\ref{fourierdec}) one must sum (only) over half-integer $n$ due to the negative signature under CPT. Solving the corresponding recursion relation with Fourier modes decaying as $a_n \sim n^{-1}$ leads to the odd spin light-rays. We have explicitly constructed the first few odd trajectories and verified that the general features discussed in figures \ref{plotofwave functions}, \ref{csfigure}, \ref{figrsurface}, \ref{figrmonodromy} are reproduced for the odd trajectories. Due to the similarity with the even spin story, we omit the details.} spin triple-twist single-trace operators. They are given by (\ref{ansatzmultitwist}) with $\Psi$ given by (\ref{rthetaansatz}) in terms of $g(\theta)$, which in turn is fixed by (\ref{recursion}) and (\ref{fourierdecay}). We can explicitly plot these kernels, see figure \ref{plotofwave functions}.



\section{Norm and structure constants} \label{normsection}

With explicit light-ray operators in hand, we can compute any correlator we want. Our next task is to relate correlators of light-ray operators to the analytic continuation of local CFT data.
This will allow us to understand the origin of the decoupling zeroes in \cite{Homrich:2022mmd}. Local CFT data is usually presented in conventions where local two-point functions have a standard normalization. Thus, in order to relate correlators of light-ray operators to this data, we must understand how to properly define the norm of light-ray operators.

%
%

This is not totally straightforward: light-ray operators annihilate the vaccuum \cite{Kravchuk:2018htv}
\beq
\mathbb{O}(X,Z) |\Omega\rangle =0
\eeq
and therefore Wightman two-point functions do not define a proper norm for these continuous spin operators.

A natural suggestion is, instead, to capture the light-rays norms through their time ordered correlators, as suggested in \cite{Caron-Huot:2013fea}. One could thus hope for a formula of the form\footnote{Such a formula was derived by Henriksson, Kravchuk and Oertel in \cite{Henriksson:2023cnh}; Our results were presented and discussed simultaneously with theirs at the Bootstrap Annual Meeting in 2023, in Sao Paulo, see talks by Kravchuk and Homrich \cite{btalks}. The derivation in \cite{Henriksson:2023cnh} is more general, being performed directly at the non-perturbative level by relying on the general construction of light-ray operators of \cite{Kravchuk:2018htv} and their relation to the inversion formula \cite{Caron-Huot:2017vep}. We believe that the perturbative argument in this section is nevertheless instructive, and we present it here for completeness. \label{petrfootnote}}
\beq
C_{\mathcal{O}_S O_1 O_2}^2(S) \stackrel{?}{=} f_{\mathcal{O}_S O_1 O_2}(S) \left(\frac{\langle\mathcal{O}_1 \mathbb{O} \mathcal{O}_2\rangle}{\langle0|\mathcal{O}_1 \mathbb{O} \mathcal{O}_2|0\rangle}\right)^2 \frac{\langle 0 | \mathbb{O} \bar{\mathbb{O}} |0\rangle}{\langle \mathcal{T}\{\mathbb{O} \bar{\mathbb{O}} \} \rangle}, \label{guessformula}
\eeq
where $f_{\mathcal{O}_S O_1 O_2}(S)$ is some universal function depending exclusively on the various conformal quantum numbers of the operators, $\langle0|\mathcal{O}_1 \mathbb{O} \mathcal{O}_2|0\rangle$ and $\langle 0 | \mathbb{O} \bar{\mathbb{O}} |0\rangle$ are canonical conformal two- and three-point structures defined in appendix \ref{perturbativeappendix}, and $\langle\mathcal{O}_1 \mathbb{O} \mathcal{O}_2\rangle$, $\langle \mathcal{T}\{\mathbb{O} \bar{\mathbb{O}} \} \rangle$ are the dynamical correlators. Let us stress that, out of the various factors in equation (\ref{guessformula}), only $\langle\mathcal{O}_1 \mathbb{O} \mathcal{O}_2\rangle^2/\langle \mathcal{T}\{\mathbb{O} \bar{\mathbb{O}} \} \rangle$ is dynamical and theory dependant, all other factors are kinematical.

The definition of a ``time-ordered'' correlator of light-ray operators is somewhat subtle. In a perturbative CFT, we will define it by starting with a time-ordered correlator of the constituent fields that make up the light-ray operator, and then integrating against the appropriate kernel $\Psi$. In a nonperturbative CFT, one approach is to construct the light-ray operator via a bilocal integral of local operators \cite{Kravchuk:2018htv}, and apply the bilocal kernel to a time-ordered correlator (as in the derivation of \cite{Henriksson:2023cnh}). The relationship between these approaches is not obvious. Fortunately, the simple perturbative definition will be suitable for this work.

\subsection{(Twist 2 motivation of the) complex $S$ structure constant formula}
To get some intuition on how the formula (\ref{guessformula}) might work, let us consider the case of twist two light-ray operators,
\beq
\!\!\!\!\!\!\!\!\mathbb{O}_2 = \frac{i \sqrt{2} \sqrt{\Gamma(2S+1)}}{2 \pi \Gamma(S+1)} \!\!\int\limits_{-\infty}^{\infty}\!\! d\alpha_1 d\alpha_2 
{\underbrace{\left(\frac{1}{(\alpha_1 - \alpha_2 + i \epsilon)^{1+S}} + \frac{1}{(\alpha_2-\alpha_1 + i \epsilon)^{1+S}} \right)}_{\psi(\alpha_1,\alpha_2)}} 
\text{Tr}\left(\mathcal{Z}(y + \alpha_1 z) \mathcal{Z}(x + \alpha_2 z)\right) \label{twist2lightray}
\eeq
at leading order. 
The full computation is 
completely fixed by conformal symmetry in this case, and thus we can explicitly keep track of all factors contributing to (\ref{guessformula}). We will take $O_1 = \Tr\left(\mathcal{X}\bar{\mathcal{Z}}\right)$ and $O_2 = \Tr\left(\bar{\mathcal{X}}\bar{\mathcal{Z}}\right)$ in this example.\footnote{That is, we compute structure constants of light-ray operators and two $1/2$-BPS operators of which $O_1$ and $O_2$ are the relevant terms.}

At even integer spin, these light-rays reduce to light-transforms of local twist-2 operators, $$\phantom{ }\qquad \mathbb{O}_S(x,z) = L[\mathcal{O_S}](x,z) \qquad \text{for even integer $S$,} $$ where
\begin{equation}
\mathcal{O}_S(x,z)=\frac{1}{\sqrt{2(2S)!}} \sum_{i}(-1)^i \binom{S}{i}^2 \text{Tr}\left( D^i_{z} \mathcal{Z} D^{S-i}_{z} \mathcal{Z}\right)\hspace{-0.1cm}(x)
\nonumber.
\end{equation}

The operators $\mathcal{O}_S$ are canonically normalized, 
\beq
\langle \mathcal{O}(x_1,z_1) \mathcal{O}(x_2,z_2) \rangle = \frac{\left((x_{12}\cdot z_1)(x_{12}\cdot z_2) - \tfrac{1}{2}x_{12}^2 (z_1 \cdot z_2 )\right)^S}{(x_{12}^2)^{\Delta + S}},
\eeq
and the three point function of interest is
\beq
\langle \mathcal{O}_1(x_1) \mathcal{O}_S(x_3,z) \mathcal{O}_2(x_2)\rangle =  \frac{\sqrt{2}S!}{\sqrt{2S!}}\frac{\left(z\cdot x_{23} x_{13}^2- z\cdot x_{13} x_{23}^2\right)^S}{x_{23}^{\Delta + S}x_{13}^{\Delta + S}x_{12}^2}, \label{local3ptcomputationtwist2}
\eeq
from which one can read that $C_{\mathcal{O}_S O_1 O_2}(S) = \sqrt{2} \Gamma(S+1)/\sqrt{\Gamma(2S+1)}$. 

Let us reproduce this result directly at complex spin from $\langle \mathcal{O}_1 \mathbb{O} \mathcal{O}_2\rangle$ and $\langle \mathcal{T}\{\mathbb{O}\bar{\mathbb{O}}\}\rangle$. At leading order these correlators are given by simple Wick contractions. The computation is discussed briefly in figure \ref{twist2compfig}. Details and further discussion can be found in appendix \ref{perturbativeappendix}. 

 The computation of $\langle \mathcal{O}_1 \mathbb{O} \mathcal{O}_2\rangle$ is straightforward. One simply deforms the $\alpha_i$ integrals and picks up the propagator poles. The result is
 \beq
\frac{\langle \mathcal{O}_1(x_1) \mathbb{O} \mathcal{O}_2(x_2)\rangle}{\langle0|\mathcal{O}_1 \mathbb{O} \mathcal{O}_2|0\rangle} = \left(-2\pi i \frac{\Gamma(2S+1)}{\Gamma(S+1)^2} \right)  C_{\mathcal{O}_S O_1 O_2}(S). \label{final3ptmaintext}
 \eeq

The term in parenthesis can be understood as resulting from the light-transform of the canonical three-point structure \cite{Kravchuk:2018htv}. Indeed, now restoring explicit dependence on the twist $\tau = \Delta-J$ and allowing for generic dimensions $\Delta_1, \Delta_2$ for the scalar operators $\mathcal{O}_1$ and $\mathcal{O}_2$, we have
\begin{align}
&\int dx_3^+\frac{(x_{2}^- x_{13}^2 - x_{1}^- x_{23}^2)^S}{x_{23}^{\tau + 2S + \Delta_2 - \Delta_1}x_{13}^{\tau + 2S + \Delta_1 - \Delta_2}x_{12}^{ \Delta_1 + \Delta_2 - \tau}} = \\&\left( \frac{-2\pi i \Gamma(\tau + 2S -1)}{\Gamma(\tfrac{\tau + \Delta_1 - \Delta_2 + 2 S}{2})\Gamma(\tfrac{\tau + \Delta_2 - \Delta_1 + 2 S}{2})}\right)
 \frac{\left(\tfrac{x_1^2}{x_1^-}-\tfrac{x_2^2}{x_2^-}\right)^{1-\tau -S}}{x_{12}^{\Delta_1 + \Delta_2 -\tau}(x_{1}^-)^{(\Delta_1 - \Delta_2 + \tau)/2}(-x_{2}^-)^{(\Delta_2 - \Delta_1 + \tau)/2}} \nonumber.
\end{align}
The second term above is our definition for the kinematical three point conformal structure $\langle0|\mathcal{O}_1(x_1) \mathbb{O}(-\infty n^+, n^+)\mathcal{O}_2(x_2)|0\rangle$. The numerical prefactor in parenthesis will be absorbed as part of $f_{\mathcal{O}_S O_1 O_2}(S)$ in (\ref{guessformula}).


\begin{figure}[t]
\centering
\includegraphics[width=\textwidth]{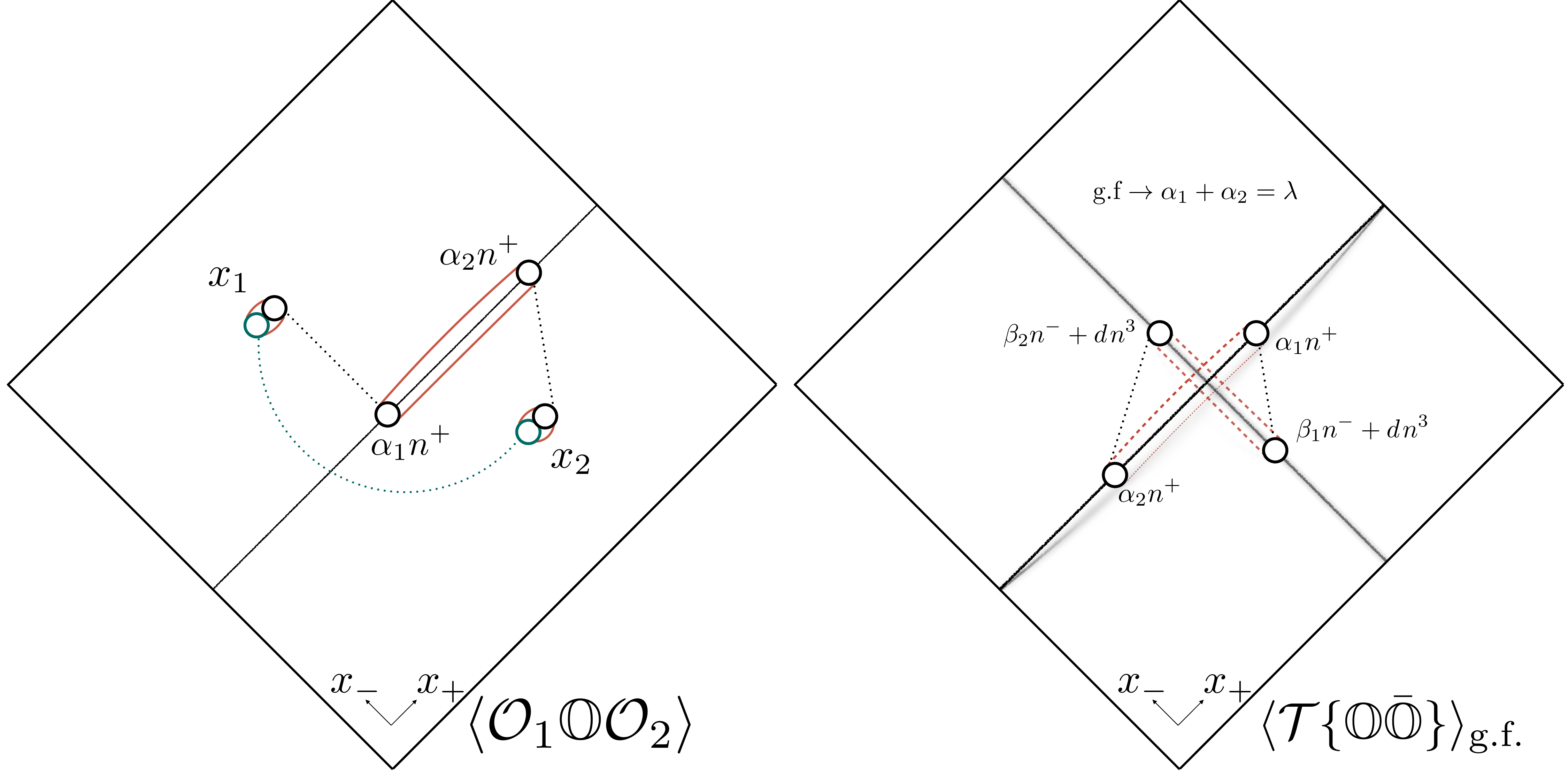}
\vspace{-0.3cm}
\caption{Structure constants between the non-local operator and two local scalar operator are obtained by Wick contracting the constituents of the line operator with the scalars (left) as well as normalizing the light-ray operator by dividing by its norm (right). Here we are considering the twist two case where two partons are integrated along a light-ray. The left picture yields (\ref{final3ptmaintext}) while the right picture leads to (\ref{finalnormmaintext}).}
\label{twist2compfig}
\end{figure}

Next we move on to the computation of $\langle \mathcal{T}\{\mathbb{O}(x_1, z_1) \bar{\mathbb{O}}(x_2, z_2) \} \rangle$. The computation is immediately more subtle. The correlator is invariant under the action of a $SO(1,1)$ conformal transformation --- the $M_{+-}$ boost in the frame of figure \ref{twist2compfig}. Its action on the fields of (\ref{twist2lightray}) can be absorbed by rescaling the integration variables. 

One thus has a zero-mode in the integrals. It produces a divergent factor $\text{vol}(SO(1,1))$ upon integration. We will divide the correlator by this factor to obtain a finite result. In practice, this means `gauge fixing' the action of the $M_{+-}$ boost and inserting the proper Faddeev-Popov determinant, which guarantees the result is gauge independent. 

We thus want to compute
\begin{equation}
\langle \mathcal{T}\{\mathbb{O}\bar{\mathbb{O}} \} \rangle_{\text{g.f.}} = \frac{\Gamma(2S+1)}{4 \pi^2 \Gamma(S+1)^2}\int 
\prod_{i=1}^2d\alpha_i d\beta_i 
\frac{\lambda \delta(\alpha_1 + \alpha_2 -\lambda) \psi(\alpha_1, \alpha_2) \psi(\beta_1, \beta_2)}{(\alpha_1 \beta_1 + d^2 + i \epsilon)(\alpha_2 \beta_2 + d^2 + i \epsilon)}, \label{startingpointtwopoint}
\end{equation}
where we inserted the gauge fixing (g.f.) condition $\alpha_1 + \alpha_2 = \lambda$, with the overall factor of $\lambda$ being the Faddeev-Popov determinant ensuring that the result is independent of $\lambda$. As in the three-point element computation, there are two possible Wick contractions which can be combined into a factor of two by permutation symmetry of the wave function $\psi(\beta_1,\beta_2)$. 

The computation is performed in appendix \ref{perturbativeappendix}. The result is 
\beq
\frac{\langle \mathcal{T}\{\mathbb{O}\bar{\mathbb{O}} \} \rangle_{\text{g.f.}}}{\langle 0 | \mathbb{O} \bar{\mathbb{O}} |0\rangle} = -\frac{ \left( 1 + e^{i\pi (-1-S)}\right)}{(2S+1)}. \label{finalnormmaintext}
\eeq

Combining the results (\ref{final3ptmaintext}) and (\ref{finalnormmaintext}) allows us to write a sharp version of (\ref{guessformula}):

\begin{center}
\begin{minipage}{0.9\textwidth}
\begin{mdframed}[style=MyFrame]
\begin{center}
    \textbf{{Continuation of Structure Constants from Light-Ray Correlators}}
    \end{center}
    \vspace{0.3cm}
    The analytic continuation in spin of structure constants is given in terms of light-ray correlators as
\beq C_{\mathcal{O}_S O_1 O_2}^2(S) = f_{\mathcal{O}_S O_1 O_2}(S) \left(\frac{\langle\mathcal{O}_1 \mathbb{O} \mathcal{O}_2\rangle}{\langle0|\mathcal{O}_1 \mathbb{O} \mathcal{O}_2|0\rangle}\right)^2 \frac{\langle 0 | \mathbb{O} \bar{\mathbb{O}} |0\rangle}{\langle \mathcal{T}\{\mathbb{O} \bar{\mathbb{O}} \} \rangle_\text{g.f.} } \label{finalformulastructure}
\eeq
with
\beq f_{\mathcal{O}_S O_1 O_2}(S) \equiv\frac{ e^{-i \pi S} - 1}{\Delta + S -1}\left(\frac{\Gamma\left(\tfrac{\tau + \Delta_1 - \Delta_2 + 2 S}{2}\right)\Gamma\left(\tfrac{\tau + \Delta_2 - \Delta_1 + 2 S}{2}\right)}{ 2 \pi i \Gamma(\tau + 2S -1)}\right)^{2} \label{conversionfactor}
\eeq

where $\langle0|\mathcal{O}_1 \mathbb{O} \mathcal{O}_2|0\rangle$ and $\langle 0 | \mathbb{O} \bar{\mathbb{O}} |0\rangle$ are canonical two- and three-point structures defined by the quantum numbers of the various operators, 
$\langle\mathcal{O}_1 \mathbb{O} \mathcal{O}_2\rangle$ is the three-point correlator and $\langle \mathcal{T}\{\mathbb{O} \bar{\mathbb{O}} \} \rangle_\text{g.f.}$ is the light-ray norm as defined above.
\end{mdframed}
\end{minipage}
\end{center}

We have derived this formula only in the case of twist two light-rays at leading order, so from this perspective (\ref{finalformulastructure}) is an ansatz. In the next section, we will use this formula to compute  structure constants for the various twist three light-ray operators constructed in section~(\ref{lightrayconstruction}) at complex spin. We can then compare --- at least in the right-half spin plane --- the results obtained with (\ref{finalformulastructure}) to the complex spin structure constants obtained using Newton's formula in \cite{Homrich:2022mmd} --- see the analogous (\ref{newtonseries}) for anomalous dimensions. The match is perfect. Given that the twist three light-rays are certainly not fixed by conformal symmetry and depend in detail on the dynamics of the theory, this gives robust evidence that (\ref{finalformulastructure}) is correct in general. This was our route to argue for (\ref{finalformulastructure}). As mentioned in footnote \ref{petrfootnote}, a direct non-perturbative argument is given in \cite{Henriksson:2023cnh}.

\subsection{Structure constants at twist 3 from light-ray wave functions}

The integrals computing twist two structure constants, being controlled by conformal symmetry, could be done analytically. For twist three, the dynamical correlators  $\langle \mathcal{O}_1 \mathbb{O}_{n,S} \mathcal{O}_2 \rangle$ and ${\langle \mathcal{T}\{\mathbb{O}_{n,S}\bar{\mathbb{O}}_{n,S} \} \rangle}$ are given by non-trivial integrals over the dynamical part of the twist three wave functions: the functions $g(\theta)$ fixed by the quantization condition (\ref{fourierdecay}). The computation is discussed in figure \ref{twist3comp} and the precise expressions are derived in appendix \ref{perturbativeappendix}. In this section, we quote the results and discuss the evaluation of the twist three structure constants directly from the light-rays constructed in section \ref{lightrayconstruction}. This will enable us to discuss the origin of the decoupling zeroes uncovered in \cite{Homrich:2022mmd} from the light-rays.

We parametrize twist three light-rays inserted at $x=-\infty n^+$, $z=n^+$ as
\beq
\mathbb{O}_{n,S} = \int d\alpha_{\text{cm}} dr d\theta \left(\psi\left(\alpha_{cm},r,\theta\right) \equiv \frac{1}{\Gamma(S_n^* -S)}\frac{g_{n,S}(\theta)}{r^{1+S}}\right) \text{Tr}\left(\mathcal{Z}(\alpha_1) \mathcal{Z}(\alpha_2) \mathcal{Z}(\alpha_3)\right),
 \eeq
with $\alpha_i$ being parametrized by $(\alpha_{cm},r,\theta)$ according to (\ref{alphachange1}-\ref{alphachange3}) as 
\begin{align}
\alpha_i &=  \alpha_{cm} + \tfrac{1}{3} r \alpha_i^\theta, \\ 
\alpha_i^\theta &= \cos(\theta - 2\pi (i-1)/3) - \cos(\theta - 2\pi (i-2)/3).
\end{align}
 We introduce the $\Gamma$ function factor for later convenience, where $S_n^* = 6 (n-1)$ is the first integer spin in which the $n$-th even trajectory has physical operators, see figure \ref{CFTdatafig}. 

\begin{figure}[t]
\centering
\includegraphics[width=\textwidth]{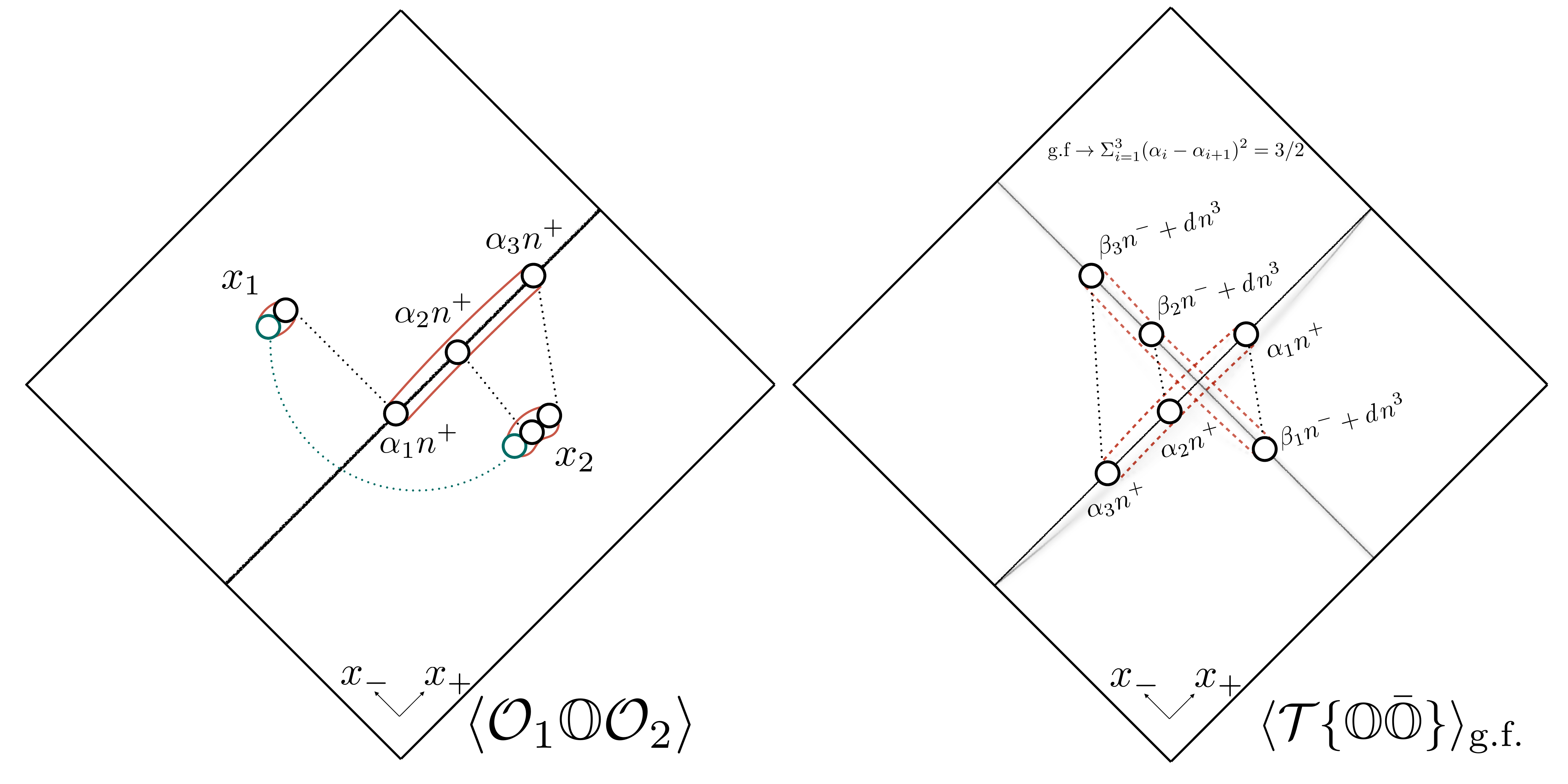}
\vspace{-0.3cm}
\caption{Same as figure \ref{twist2compfig} but for twist three operators, made of three partons. The left picture leads to (\ref{maintextfinalthreepoint}) -- once two of the three parton integrations is carried out explicitly -- while the right picture yields (\ref{finalmaintextnorm}) -- once we take care of three of the six integrations depicted in the figure.
}
\label{twist3comp}
\end{figure}

The computation of the matrix elements $\langle \mathcal{O}_1 \mathbb{O}_{n,S} \mathcal{O}_2 \rangle $ is straightforward. The triple integral is reduced to a single integral by deforming two of the integral contours, picking up propagator poles. The result is 
\begin{equation}
\frac{\langle\mathcal{O}_1 \mathbb{O} \mathcal{O}_2\rangle}{\langle0|\mathcal{O}_1 \mathbb{O} \mathcal{O}_2|0\rangle} =  \frac{1}{ \sin(\pi S) \Gamma(S^*_n - S)}
 \int  d\theta g(\theta) F_{\text{three points}}(\theta), \label{maintextfinalthreepoint}
\end{equation}
which must be evaluated numerically for each trajectory as we will discuss later. The kernel $F_{\text{three points}}$ is purely kinematical, and is defined in equation (\ref{kernelofmaintext3pt}).

As in the twist two case, the twist three norm $\langle\mathcal{T}\{\mathbb{O}\bar{\mathbb{O}}\}\rangle$ requires gauge fixing. We use the gauge $ \delta(r_\alpha - 1)$. The starting point is given by a six-dimensional integral. Two integrals can be performed by residues, while one is localized by the delta function, see appendix \ref{perturbativeappendix}. The result is 

\begin{equation}
\frac{\langle\mathcal{T}\{\mathbb{O}\bar{\mathbb{O}}\}\rangle_\text{g.f.}}{\langle 0 | \mathbb{O} \bar{\mathbb{O}} |0\rangle} = \frac{1}{\sin(\pi S) \Gamma(S_n^* - S)^2} \int_D d\alpha_{cm} d\theta_\alpha d\theta_\beta g_{\mathbb{O}}(\theta_\alpha)g_{\bar{\mathbb{O}}}(\theta_\beta) F_\text{two point}(\alpha_{cm},\theta_{\alpha} ,\theta_{\beta}),\label{finalmaintextnorm}
\end{equation}
where the kinematical kernel $F_\text{two point}$ is defined in (\ref{twopointintegral}). The integral over $\alpha_{cm}$ can be performed analytically. Instead, it is enough for our purposes to proceed numerically from this point. 

All ingredients necessary to compute the structure constants $C_{\mathbb{O}_{n,S} \mathcal{O}_1 \mathcal{O}_2}$ are now in place. \begin{enumerate}
    \item  We generate the complex spin wave functions for the various discrete trajectories following section \ref{lightrayconstruction}. The wave function $g(\theta)$ is approximated by its Fourier series truncated to some large Fourier mode. 
    \item The integrals (\ref{finaltheta}) and (\ref{twopointintegral}) are then evaluated numerically. The multidimensional integral can be straightforwardly computed through Quasi-Monte Carlo.
\end{enumerate}

\begin{figure}[t]
\centering
\includegraphics[width=\textwidth]{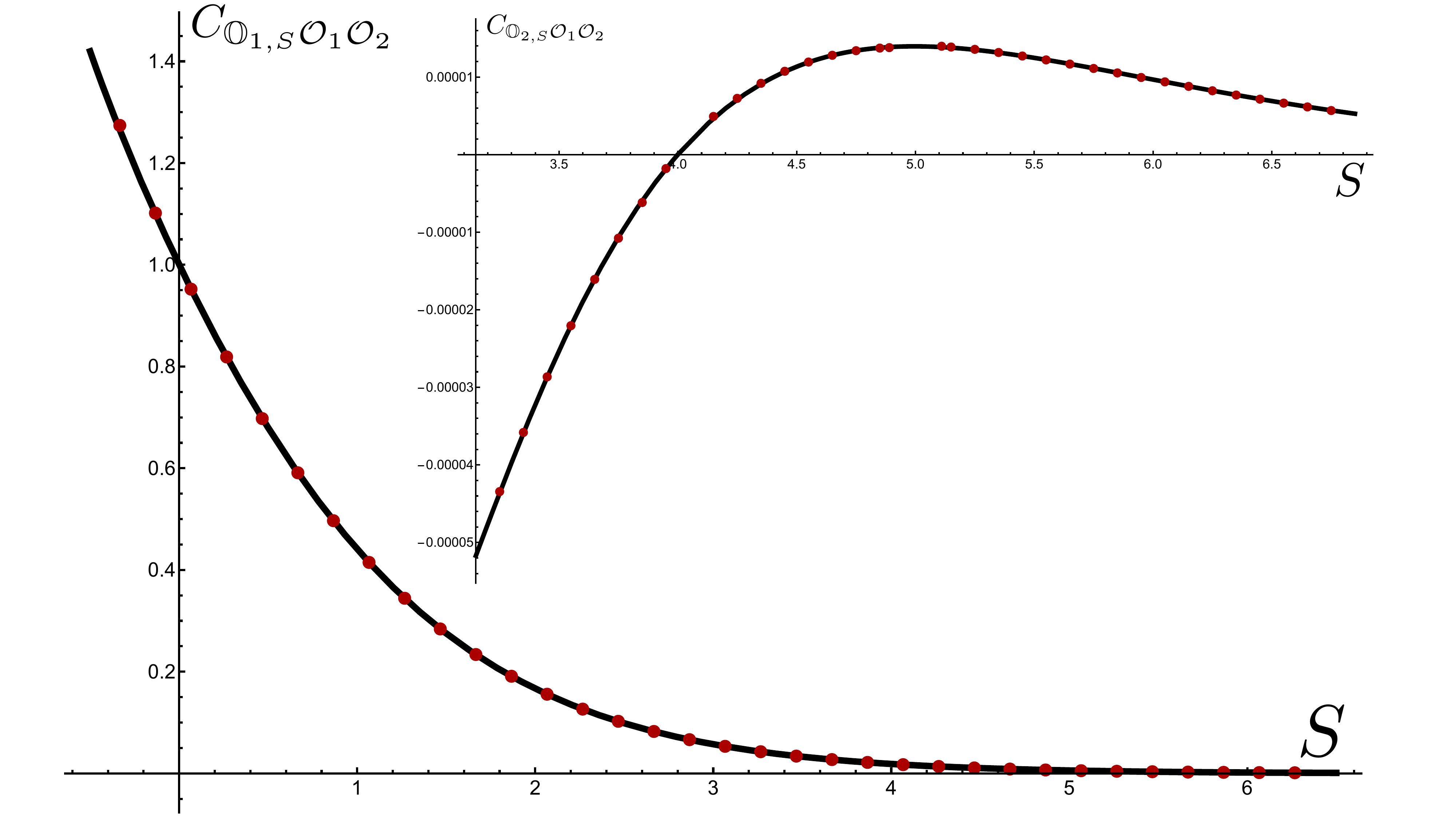}
\vspace{-0.3cm}
\caption{In red, structure constants computed from explicit light-ray operators using (\ref{finalformulastructure}). In black, the Newton interpolation (\ref{newtonseries}) for the structure constants, see \cite{Homrich:2022mmd}. They match perfectly. We plot the results for the first and second trajectory, including the first decoupling zero.}
\label{csfigure}
\end{figure}

The result of this computation is shown in figure \ref{csfigure}. There is a perfect match with the result obtained from Newton's formula (\ref{newtonseries}). In particular, the decoupling zeroes of \cite{Homrich:2022mmd} are reproduced. 

It is interesting to understand how these decoupling zeroes arise from the point of view of (\ref{finalformulastructure}). It is important to clarify that the answer to such a question depends on how one normalizes the operators. We chose to normalize the operators $\mathbb{O}_{n,S}$ through the condition $\a_m = \tfrac{1}{m} +  O(\frac{1}{m^2})$ for all values of $S$. In this case what happens to the light-ray operators as one approaches a positive even spin value $2k$ is:

\begin{itemize}
    \item For $S\rightarrow 2k$ with $2k<6(n-1)$, the Fourier modes $a_m$ with $|m|\leq \lfloor2k/6\rfloor$ vanish linearly as $(S-2k)$, other modes remaining finite and non-zero. For $S\rightarrow 2k$ with $2k\geq  6(n-1)$, all Fourier modes $a_m$ remain finite and non-zero. We introduce $S_n^* \equiv 6(n-1)$ for this important threshold value separating these two behaviors. 
    \item Meanwhile, the radial part of the wave function develops a simple pole for $S\rightarrow 2k$ with $2k<S_n^*$,
    \beq
    \frac{r^{-1-S}}{\Gamma(S_n^*-S)} = -\frac{\delta^{2k}(r)}{\Gamma(S_n^* - 2k)\Gamma(2k+1)(S-2k)} + O\left((S-2k)^0\right)\la{case1}
    \eeq
    and remains finite for $2k\geq S_n^*$:
    \beq
    \frac{r^{-1-S}}{\Gamma(S_n^*-S)} =     \frac{\Gamma(2k - S_n^*+1)\delta^{2k}(r)}{\Gamma(2k+1)} + O\left(S-2k\right) \la{case2}.
    \eeq
\end{itemize}
Thus, with this choice of normalization, we see that, for any even integer $2k$, the light-ray operators and their matrix elements such as $\langle \mathcal{O}_1 \mathbb{O}_{n,S} \mathcal{O}_2 \rangle $ remain finite and generically non-zero. 

When analyzing $\langle \mathcal{T}\{\mathbb{O} \bar{\mathbb{O}} \} \rangle_\text{g.f.}$ one needs to be more careful. The presence of the gauge-fixing factor $\delta(r_\alpha-1)$ in $\langle \mathcal{T}\{\mathbb{O} \bar{\mathbb{O}} \} \rangle_\text{g.f.}$ effectively means we should replace $r_\alpha^{-1-S} \rightarrow 1$ in the $\psi_\mathbb{O}(\alpha_i)$ wave function contributing to the light-ray norm. The absence of the $r_\alpha^{-1-S}$ factor, which previously produced a pole and a delta function has two effective consequences: first, the light-ray no longer localizes to the first few fourier modes and, second, there is no longer a pole from the distributional limit of $r^{-1-S}$ at even $S$. Thus the radial part for the gauge-fixed operator contributes an extra zero relative to (\ref{case1}) and (\ref{case2}), while the angular part no longer contributes a zero for $2k<S_n^*$. In total, we conclude that  $\langle \mathcal{T}\{\mathbb{O} \bar{\mathbb{O}} \} \rangle_\text{g.f.}$ vanishes for $2k\geq S_n^*$ and remains finite for $2k < S_n^*$.

Thus in this normalization the decoupling zeroes simply follow from $1-e^{-i \pi S}$ factor in the conversion factor~(\ref{conversionfactor}). More precisely, for even spin $S=2k \ge S_n^*$ in which local operators \textit{do} exist, the conversion factor zero cancels with the zero from the light-ray norm, resulting in a non-zero structure constant. For even spins $S=2k <S_n^*$ when the local operators do \textit{not} exist the conversion factor zero has no pole to cancel with and we get a decoupling zero for the structure constant. We summarize the various factors in the following table.
\begin{table}[h]
\begin{center}
\begin{tabular}{ |c|c|c|c|c| } 
 \hline
  & $\langle \mathcal{O}_1 \mathbb{O}_{n,S} \mathcal{O}_2\rangle$ & $\left(\langle \mathcal{T}\{\mathbb{O} \bar{\mathbb{O}} \} \rangle_\text{g.f.}\right)^{-1} $ & $f_{\mathcal{O}_S \mathcal{O}_1 \mathcal{O}_2}(S)$ & $C_{\mathcal{O}_S O_1 O_2}^2(S)$\\
 \hline
 $2k < S_n^*$ & $(S-2k)^0$ & $(S-2k)^0$ & $(S-2k)$ & $(S-2k)$\\ 
 \hline
 $2k \geq S_n^*$ & $(S-2k)^0$ & $(S-2k)^{-1}$ & $(S-2k)$ & $(S-2k)^0$ \\ 
 \hline
\end{tabular}
\end{center}
\end{table}

Of course, one could chose to normalize the operators so that 
\beq
f_{\mathcal{O}_S \mathcal{O}_1 \mathcal{O}_2}(S) \frac{\langle 0 | \mathbb{O} \bar{\mathbb{O}} |0\rangle}{\langle \mathcal{T}\{\mathbb{O} \bar{\mathbb{O}} \} \rangle_\text{g.f.} } = 1
\nonumber
\eeq
in analogy to what is conventionally done for local operators. In this case, the decoupling zeroes would come directly from the three point correlators $\langle \mathcal{O}_1 \mathbb{O}_{n,S} \mathcal{O}_2\rangle$. Both options are fine.

This was the right-half plane story from the point of view of operators. It is time to see what dragons await us in the left-half plane.

\section{The complex plane} \label{riemannsurfacesec}
\subsection{Riemann surface, expectation and result}

A main point of this paper so far has been that the various twist three operators organize themselves into infinitely many well defined Regge trajectories. They could be numbered and ordered according to their anomalous dimensions $\gamma_j$, or their $q_j$ integrable charges, continuously in spin. This picture is correct for the right-half spin plane, where local operators live and where life is plain and simple.

\begin{figure}[t!]
\centering
\includegraphics[width=\textwidth]{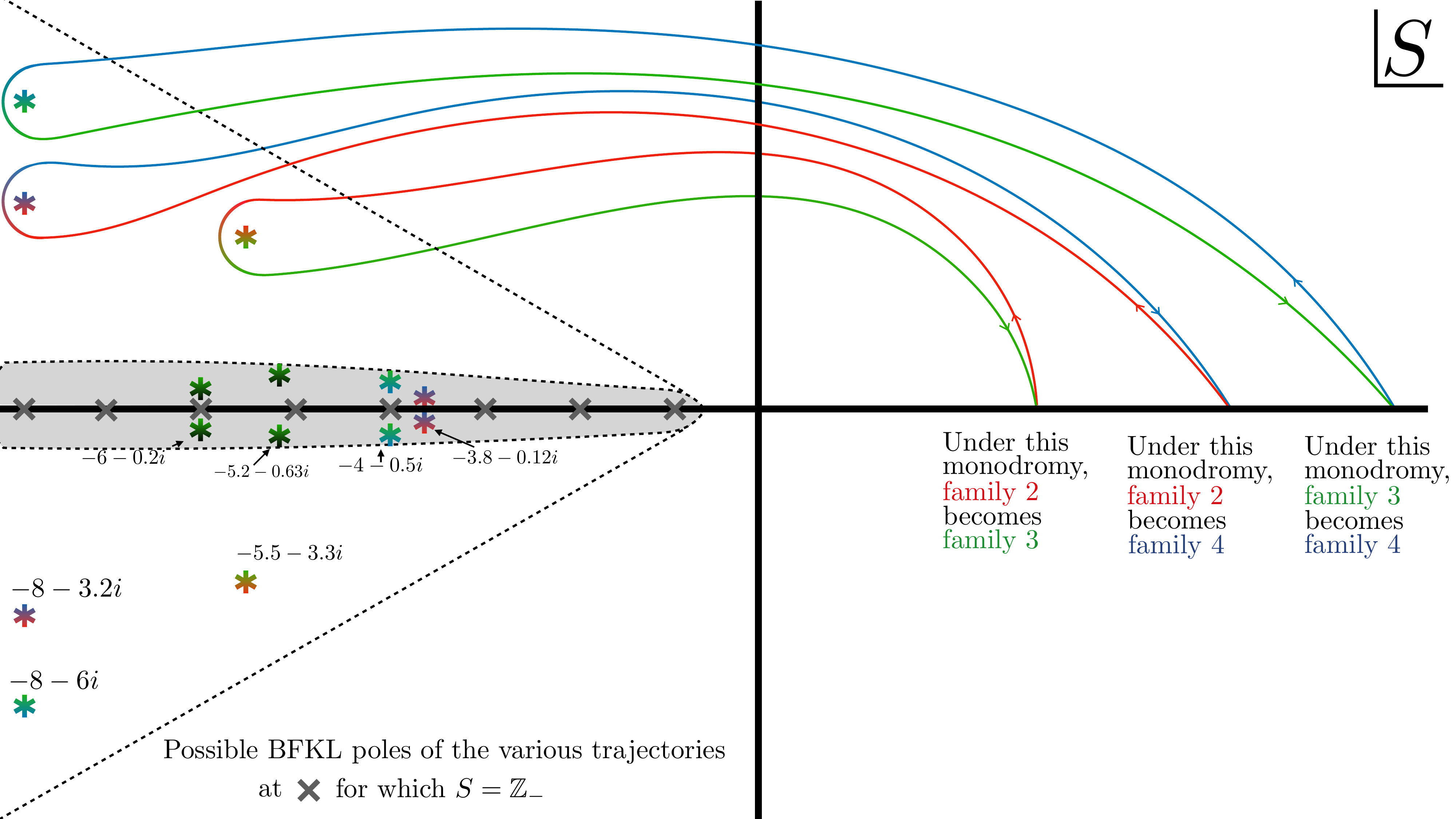}
\vspace{-0.3cm}
\caption{Here we present the analytic continuation of the various operators to the left half-plane. We observe no interesting structure to the right of a dashed cone, which includes the positive integers where the light-rays reduce to light-transforms of local operators, and where we observe the decoupling zeroes discussed in section \ref{normsection}. If we start with a certain family on the right, go for a walk on the left and come back to the same point we sometimes encounter non-trivial monodromies ending up in a different family! We zoom in one such monodromy in figure \ref{figrmonodromy}. In other words, the various even families form a single Riemann surface (except for the lowest $q=0$ trajectory which has a life of its own, see footnote \ref{lowesttfoonote}). The branch points are located inside the dashed cone and come in conjugate pairs. For $\text{Re}(S)>-8$, $\text{Im}(S)>1$ the three branch points displayed here exhaust the monodromies involving even spin families 2, 3 and 4. In the grey region, close to the real axis, we observe a high density of branch points, which in turn slow down our numerical algorithm. In this region we expect several other branch points, in particular mixing to higher trajectories, such as in the green-black branch points. There are certainly many more that we did not manage to resolve. There could even be tiny real cuts there. In that region besides these cuts we have infinitely many poles at negative integers corresponding to the mixing with horizontal trajectories. 
This figure is a quantitative depiction of the qualitative cartoon anticipated in figure 3 of \cite{Homrich:2022mmd}. The position of the branch points seem generic and we provide approximate (but rather precise) values here. All of them have ramification index 2.}
\label{figrsurface}
\end{figure}

\begin{figure}[t]
\centering
\includegraphics[width=\textwidth]{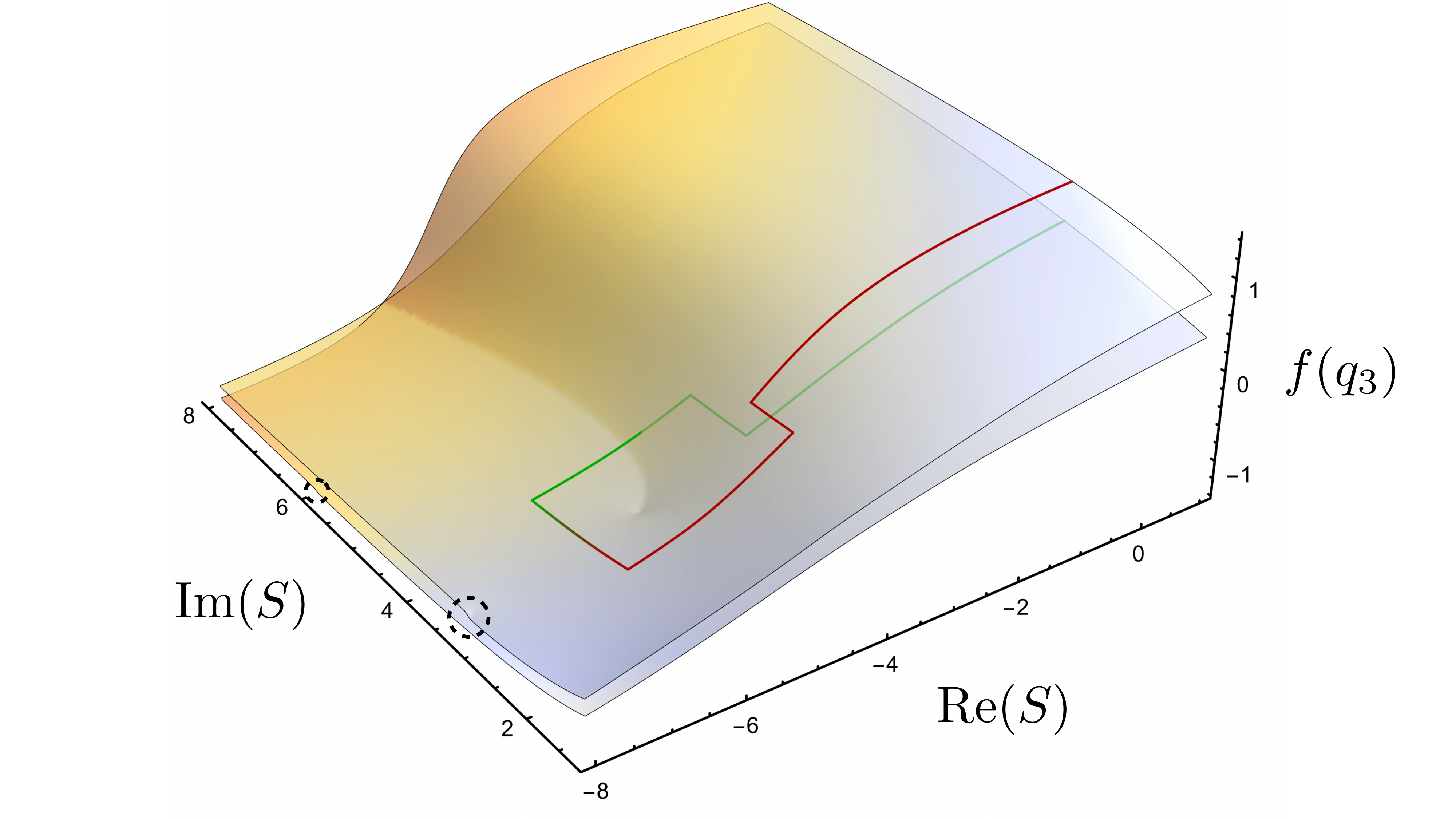}
\vspace{-0.3cm}
\caption{As we encircle the branch points around $S\simeq -5.5\pm 3.3i$ (see figure \ref{figrsurface}) families $2$ and $3$ get swapped. Here we plotted the integrability charge $q$ (or rather $f(q)= \arctan\left(\text{Im}(q)/20\right)$ which leads to a prettier plot) around that branch point to illustrate this. Two sheets of this function meet at that branch point. Of course, since the operators themselves are getting swapped, any quantity that probes them -- such as the anomalous dimensions, the structure constants or any other correlator involving them -- would produce a similar picture. We used a grid of $1.7 \times 10^6$ points to generate this plot. 
Around $\text{Re}(S) = -8$ we encircle two other branch points which connect to a third sheet which is not represented here to ease visualization. Crossing these branch cuts with similar monodromies lead to the fourth even spin trajectory, see figure \ref{figrsurface}.}
\label{figrmonodromy}
\end{figure}

The global picture is quite different: all these various well-defined trajectories merge into a single\footnote{To be precise, the lowest even trajectory remains unmixed perturbatively due to its distinct parity charge, see footnote \ref{lowesttfoonote}. At finite $\lambda$, it mixes with other parity singlets, see \cite{Klabbers:2023zdz}. Higher even and odd spin trajectories also each belong to disconnected Riemann surfaces.} analytic entity, connected through infinitely many branch points at (seemingly) generic locations in the left half spin plane, see figures \ref{figrsurface} and \ref{figrmonodromy}. Perhaps surprisingly, this happens \textit{in perturbation theory}!  This is in contrast to more familiar operator mixing such as the well studied BFKL-DGLAP-Shadow mixing \cite{Balitsky:2013npa, Brower:2006ea, Jaroszewicz:1982gr, Lipatov:1996ts,Kotikov:2000pm,Kotikov:2002ab,Caron-Huot:2022eqs}, for which branch points can only be accessed at finite coupling.\footnote{In $\mathcal{N}=4$ SYM the Riemann surface realizing this mixing has been constructed at finite coupling for the twist $2$ trajectory in \cite{Gromov:2015wca}, and recently in \cite{Klabbers:2023zdz} for the case of the leading twist three trajectory. See also \cite{Caron-Huot:2022eqs} for the direct mixing of the 45 degrees trajectory and its shadow in the Wilson-Fisher theory.} In perturbation theory the BFKL-DGLAP cuts close, and manifest as poles at negative integer values of spin.  Here instead we consider a infinite number of trajectories which are degenerate in perturbation theory, and thus mix right away, allowing us to explore their rich and complicated Riemann surface already at leading order.

An integrabilist could have foreseen this by examining the local operator data. For example, at spin $S=12$ there are three local operators whose structure constants $C_{n,S}\equiv C_{\mathcal{O}_{n,S} \mathcal{O}_1 \mathcal{O}_2}^2$ are given by
\begin{align*}
C_{1,12}^2 &=  \frac{11}{12086676},\\
C_{2,12}^2 &= \frac{206}{16034270715} + \frac{695219}{(80171353575 \sqrt{455745})},\\
C_{3,12}^2 &= \frac{206}{16034270715} - \frac{695219}{(80171353575 \sqrt{455745})}.
\end{align*}
with a similar multi-square-root structure being present at higher values of spin. These strongly suggest the various families (with the exception of the lowest trajectory which remains rational at all value of spin) can be analytically continued into each other. 

 In section \ref{howderivedsurface} we will describe how to reconstruct the twist three Riemann surface numerically using integrability methods. The results are described in figures \ref{figrsurface} and \ref{figrmonodromy}. In section \ref{couldbederived} we discuss an alternative strategy from which one could in principle reconstruct the Riemann surface from the local operator spectrum.

\subsection{How it is derived: integrability}
\label{howderivedsurface}
To generate energies at complex spin and construct the rich Riemann surfaces and associated monodromies, we diagonalize the Baxter operator, whose action on light-rays is discussed in appendix \ref{baxterappendix}. In practice, we solve the twist-three Baxter equation
\beq
Q(u+i) (u+i/2)^3+Q(u-i) (u-i/2)^3 = Q(u) \times (\underbrace{2u^3-(S^2+2S+3/2)u+q}_{t(u)}) \,, \label{Baxter} 
\eeq
with $S$ complex. One algorithm for solving this equation was suggested in \cite{Homrich:2022mmd}; it works perfectly for $S$ in the right half-plane but it does not work for $\text{Re}(S)<-1$. We suggest here an alternative algorithm very close in spirit to \cite{Gromov:2015wca} which works for any $S$ and agrees with the one in \cite{Homrich:2022mmd} for $\text{Re}(S)>-1$.\footnote{We thank Kolya Gromov for discussions which lead to the algorithm we discuss below. Simon Caron-Huot recently presented yet another numerical algorithm to solve the $L=3$ Baxter equations at complex spin during a lecture series at ICTP-SAIFR \cite{simonlectures}. We thank Simon for several discussions.} The algorithm is based on three simple steps:
\begin{itemize}
\item \textit{For any} $q$, there exist two special solutions to the Baxter equation with pure power asymptotics. We define them through their behavior as we go \textit{up}, $u \to +i \infty$, 
\begin{align}
Q_{-2-S}(u)&=u^{-2-S}+a_1 u^{-3-S}+a_2 u^{-4-S}+\dots\,,\nonumber\\
 Q_{S}(u)& =u^{S}+b_1 u^{S-1}+b_2 u^{S-2}+\dots\,.
\end{align}
Importantly, note that the leading power is canonically normalized to $1$, and furthermore note that plugging such expansions into the Baxter equation allows us to fix completely all $a_j$ and $b_j$. For example, 
\beq
a_1= -\frac{q}{2 S+3}, \qquad a_2=\frac{q^2}{4 (S+2) (2 S+3)}+\frac{1}{48} (S+2) \left(S^2+4 S+6\right)\,, \qquad \text{etc}.
\eeq
So as a first step in the algorithm we compute the functions $\hat{Q}_{-2-S}$ and $\hat{Q}_{S}$ given as a large $u$ expansion truncated after $N$ terms $\{a_1,\dots,a_N\}$ and $\{b_1,\dots,b_N\}$. For $N$ large, such functions $\hat{Q}_{-2-S}$ and $\hat{Q}_{S}$  are excellent approximations to ${Q}_{-2-S}$ and ${Q}_{S}$ as long as $u$ has a big imaginary part. How big should the imaginary part be and for a given large cut-off $N$ can be found experimentally checking for proper convergence.  
\item The functions $Q_S$ and $Q_{-2-S}$ are holomorphic in the upper half plane but they are singular in the lower half plane. Indeed, suppose we have a solution to the Baxter equation in the upper half place, valid for for some large imaginary part $u$. Using Baxter repeatedly we can lower the imaginary part in integer steps. Explicitly we have
\beq
Q(u)  =\frac{t(u+i) Q(u+i)-Q(u+2i) (u+3i/2)^3}{(u+i/2)^3} \,. \label{BaxterLower} 
\eeq
so that the right hand side will develop a third order pole at $u=-i/2$ unless the numerator is carefully tuned.
This pole at $u=-i/2$ then propagates to $-3i/2$ and then to $-5i/2$ etc. To construct fully holomorphic solutions to Baxter thus requires tuning and --- in particular --- it requires dropping pure power asymptotics. Instead we consider 
\beq
Q_\text{holomorphic}(u) \equiv (e^{2\pi u} + c_0 + c_1 e^{-2\pi u}) \,Q_{-2-S}(u)+ c_2 Q_S(u),   \la{asympbaxter}
\eeq
which of course solves Baxter as long and $Q_{-2-S}$ and $Q_S$ do.
Then we tune the three constants $c_0,c_1,c_2$ to require that the third-, second- and first-order pole at $u=-i/2$ vanishes. Once the function is regular at this point, it will no longer generate poles at $-3i/2, -5i/2,\dots$ and will thus be holomorphic everywhere. At infinity it behaves as~$e^{2\pi u} u^{-2-S}$ which we assume is the correct behavior for the $Q$-functions.\footnote{Similar asymptotics were advocated in \cite{Janik:2013nqa} to be the correct ones when dealing with complex spin for twist two operators. In \cite{Homrich:2022mmd} we argued for this solution to the be the proper one for twist three as well. For~$\text{Re}(S)>-1$, $Q_\text{holomorphic}$ this is the slowest growing \textit{holomorphic} solution to Baxter equation, a fact which was implicitly used in the algorithm in \cite{Homrich:2022mmd} and which is is \textit{not} being used in the new argument presented here.}

In practice we use the approximate solutions $\hat Q_{-2-S}$ and $\hat Q_S$ to obtain in this way an approximate solution  $\hat Q_\text{holomorphic}$.
\item
Note that up to now we have constructed (approximate) holomorphic solutions for any complex spin $S$ and complex charge $q$. Lastly, we quantize $q$ by imposing the zero momentum condition 
\beq
0=Q_\text{holomorphic}(i/2)-Q_\text{holomorphic}(-i/2) \la{momCond} \,. 
\eeq
Numerically solving this equation, e.g.\ with Newton's method, will lead to many $q$'s for each complex spin $S$. The various solutions are the various Regge trajectories! 

In practice we change $S$ adiabatically. We find some $q$ corresponding to the $n$-th Regge trajectory for some complex spin $S$ and then we change $S$ slightly and use that previous value of $q$ as a starting point for a new scan of (\ref{momCond}). In this way we can follow $Q$ (and thus the corresponding charges including the energy) as we follow arbitrarily complicated paths in the complex $S$ plane!  Sometimes we find that under closed paths $Q$'s acquire monodromies and do not come back to themselves. This is fine; it simply means they live on a Riemann surface and that different Regge trajectories are actually different branches of the very same master function. As long as our complex spin $S$ paths do not get too close to the branch points of this Riemann surface, the numerics will be fast and stable.

\item Finally, to extract energies, one can consider the expansion of the quantized $Q_{\text{holomorphic}}$ at $u \pm i/2$,
\beq
\gamma(S) = 2 \left(Q'(-i/2) - Q'(i/2)\right)/Q(i/2).
\eeq

\end{itemize}

Note that the Baxter equation (\ref{Baxter}), the holomorphicity condition, the zero momenta condition (\ref{momCond}) and the asymptotics (\ref{asympbaxter}) are not new conditions but instead should \textit{follow} from the action of the Baxter operator on the light-rays constructed in section \ref{lightrayconstruction}. Indeed, we discuss how to derive all these properties in appendix \ref{baxterappendix}, with the exception of the asymptotics which we only managed to derive for $L=2$. In particular, we discuss how, in the light-ray language, one can simply derive Janik's $Q$-function from the twist two light-ray wave function (\ref{twist2lightray}).

\subsection{How it could in principle have been derived: dispersion relation}
\label{couldbederived}
Could we have derived such nice Riemann surfaces with all their various sheets from a clever analytic continuation of the integer spin data, in the spirit of Newton's method employed in \cite{Homrich:2022mmd} without any of the fancier methods just developed? 

In principle \textit{Yes}.\footnote{In practice \textit{no way}. \label{noWay}}

\begin{figure}[t]
\centering
\includegraphics[width=\textwidth]{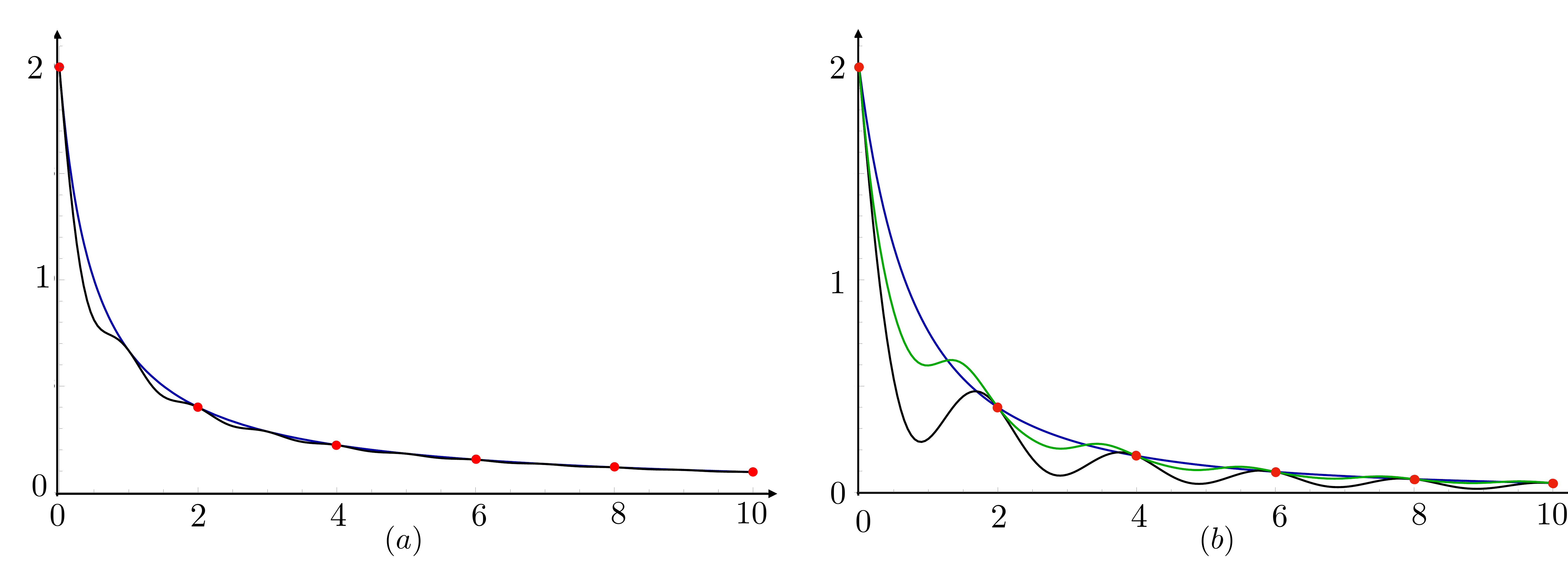}
\caption{(a) Blue: $2/(2S +1)$. Black: $ \int_{-1}^1 P_S(x)^2 dx$ which can be evaluated with Mathematica's continuation of the Legendre polynomials. They agree at the integers. (b) Blue: non-exponential analytic continuation from the integers as described in this section. Black: continuation of the sum in equation (6) of \cite{Homrich:2022mmd}. Green: $ \int_{-1}^1 P_S(x)^4 dx$ from Mathematica.}
\label{p2fig}
\end{figure}

In this section we explain how. As a warm up suppose we want to take a trivial integral that we know how to compute very well for integer spin, namely, 
\beq
I_2(S) \equiv \frac{1}{2}\int\limits_{-1}^{+1}dz\, (P_S(z))^2 = \frac{1}{2S+1}. \la{true2}
\eeq
Suppose we are given (\ref{true2}) for $S=0,1,2,3,\dots$. How can we analytically continue this integral away from the integers to reproduce the expected result $1/(2S+1)$ for any complex spin? What we should \textit{not} do is use \texttt{Mathematica} blindly: \verb"NIntegrate[LegendreP[S,z]^2,{z,-1,1}]/2" does \textit{not} give $1/(2S+1)$ unless $S$ is integer. 
Why do we like $1/(2S+1)$ as an continuation to complex spin compared to what Mathematica gives and which we plot in figure \ref{p2fig}? 

The key difference is that $1/(2S+1)$ does not blow up exponentially for large complex spin. By Carlson's theorem, it is the unique analytic continuation of the integer sequence $\{1/1,1/3,1/5,1/7,1/9,\dots\}$ with this property. Mathematica's continuation does blow up. This can be seen from the asymptotics of $\verb"LegendreP[S,z]"$, which read for $|z| \leq 1$
\beq
{P_S(-z)} \sim \frac{1}{\sqrt{S}} \left(e^{i S \theta} + e^{-i S \theta}\right), \qquad z= \cos \theta,
\eeq
which diverges with exponential type $\pi$ in the right-half plane (violating Carlsons assumptions). The oscillations in figure \ref{p2fig} are actually a sign of such $e^{2\pi iS}$ like terms which oscillate when $S$ is real but explode for complex $S$. Of course, it is easy to cook up combinations of such oscillations as in $\sin(\pi S)$ which vanish for all the integers which is why Mathematica's continuation passes through all the correct black dots in the figure despite being \textit{wrong} as long as we are after the non-exploding unique continuation.  

Let's then set Mathematica aside and try to be cleverer in our analytic continuation. We want to start with integer $S$ and do some massaging to arrive at an expression where there are no $S$ oscillations as we continue to complex $S$. The trick is to do a few cleaver contour manipulations that will end up transforming an oscillating contour integral into an exponentially damped integral over a simple discontinuity. We were inspired from the same type of manipulations performed in Regge theory to make sense of partial waves at complex spin, see \cite{Caron-Huot:2017vep}.

We start with
\beq
\frac{1}{2} \int\limits_{-1}^1 dx \, P_S(x)^2, 
\eeq
which for integer $S$ is obviously equal to 
\beq
\oint \frac{dt}{2\pi i} \frac{1}{t^{S+1}} F(t) \qquad \text{where}  \qquad F(t)=\sum_{S'=0}^\infty  \frac{t^{S'}}{2} \int\limits_{-1}^1 dx \, P_{S'}(x)^2    \label{tint} 
\eeq
Now note that to compute $F(t)$ all we need to do it add up integer spin $S'$ for which we do know (\ref{true2}). We obtain that $F(t)$ is a nice holomorphic function in the plane minus a single cut for $t>1$, 
\beq
F(t)= \sum_{S'=0}^\infty  \frac{t^{S'}}{2S'+1}  = \frac{1}{2\sqrt{t}} \log \left(\frac{1+\sqrt{t}}{1-\sqrt{t}}\right). \la{logF}
\eeq
Now we blow the contour integral in (\ref{tint}) so it becomes 
\beq
\int_{1}^\infty \frac{dt}{t^{S+1}} \frac{\text{Disc}(F)(t)}{2\pi i}.  \la{discT}
\eeq
The discontinuity of (\ref{logF}) is trivial so we end up simply with 
\beq
\frac{1}{2}\int_{1}^\infty \frac{dt}{t^{S+3/2}} = \frac{1}{2S+1},
\eeq
as we wanted. The key was going from (\ref{tint}) to (\ref{discT}). In this second representation the integrand is manifestly damped. As such this representation should now be free from oscillations and thus ought to yield the unique continuation a la Carlson theorem. And it did indeed.

This was too trivial of an example. Lets try a much more interesting integral where the values at the first few integer values of $S$ are by no means enough to guess the general result, 
\beqa
I_3(S)\equiv\frac{1}{2} \int\limits_{-1}^{+1} dx (P_S(x))^4 =  \left\{1,\frac{1}{5},\frac{3}{35},\frac{241}{5005},\frac{529}{17017},\dots\right\} \quad  \, \quad S=0,1,2,3,4,\dots\,.  \label{i3}
\eeqa
This is not a random example. Structure constants of the lowest twist three trajectory are given by the this integral up to trivial combinatorial factors.\footnote{Indeed, to recover the infinite sum in \cite{Homrich:2022mmd} from (\ref{i3}) for the structure constant of the leading trajectory, we recall the product of angular momentum identity 
\beq
P_S(x)^2 = \sum_{l=0}^\infty (4l+1) \(\medspace 
\begin{array}{ccc}
S & S & 2l \\
0 & 0 & 0 \\
\end{array}
\medspace \)^2 P_{2l}(x)
\eeq where $\(\medspace 
\begin{array}{ccc}
 j_1 & j_2 & j_3 \\
 m_1 & m_2 & m_3 \\
\end{array}
\medspace \)$ are the Wigner-3j symbols given by \texttt{ThreeJSymbol} in Mathematica. Therefore, using the orthogonality of Legendre's,
\beq
\frac{1}{2}\int P_S(x)^4 =\frac{1}{2}\int (P_S(x)^2 )^2 = \sum_{l=0}^{\infty} \frac{1}{2\times 2l+1} \left( (4l+1) \(\medspace 
\begin{array}{ccc}
S & S & 2l \\
0 & 0 & 0 \\
\end{array}
\medspace \)^2 
  \right)^2
\eeq
which is nothing but the infinite sum in \cite{Homrich:2022mmd} once we evaluate the 3J symbols in terms of a bunch of gamma functions.  
} In \cite{Homrich:2022mmd} we dealt with the analytic continuation of this integral by Newton Method; here we will instead use a numerical implementation of the method employed above. The match between the two methods is perfect as depicted in figure \ref{p2fig}b.

\begin{enumerate}
\item First we construct 
\beq
\tilde{F}(t) \equiv \sum_{S'=0}^{S_\text{max}} I_3(S') t^{S'}, \la{tildeF}
\eeq
This function will have a cut at $t=1$ as in our analytic example (\ref{logF}). However, we can not simply go to $t\pm i0$ with $t>1$ to see this log since the Taylor expansion in $t$ will only converge in a radius $1$ circle around the origin. 
\item To solve this limitation we introduce the uniformizing variable
\beq
\rho(t) = \frac{\sqrt{1-t} - 1}{\sqrt{1-t}+1},
\eeq
which maps the cut plane  $\mathbb{C}-(1,\infty)$ to the unit disk and re-expand (\ref{tildeF}) to define 
 \beq
 \tilde{\tilde{F}}(\rho) = \sum_{S=0}^{S_{\text{max}}} c_S \rho^{S}.
 \eeq
This function with two tildes is fixed to have the same expansion around $t=0$ (corresponding to the center of the unit disk) as (\ref{tildeF}) but it now converges everywhere in the $\rho$ unit disk! This allows us to access any $t$ in the cut plane!
\item Finally, we can now estimate the discontinuity of ${F}(t) \equiv \sum_{S'=0}^{\infty} I_3(S') t^{S'}$ for $t>1$ as 
\beq
{F}(t+i0)-{F}(t-i0) = \tilde{\tilde{F}}(\rho) -\tilde{\tilde{F}}(1/\rho), \label{discap}
\eeq
with $\rho$ in the upper arc of the unit disk.
\item Assuming we can blow the contour up in the general case, the desired integral can then be approximated by inserting  (\ref{discap}) in (\ref{discT}) for $z>1$. We then recover the smooth dark blue curve in (b) of figure \ref{p2fig}, reproducing the continuations of the structure constants obtained in \cite{Homrich:2022mmd} using Newton's series. 
\end{enumerate}

One can also successfully reproduce the Newton series continuation of the higher twist three families through this procedure. This provides a new perspective on the decoupling zeroes of \cite{Homrich:2022mmd}. We define, in the even twist three case,
\beq
F(t) = \sum_{n=S^*_k/2}^{\infty} \left(f_{(S\equiv2n),k}\equiv\frac{\Gamma(2n+1)^2}{\Gamma(4n+2) C_{2n,k}^2 }\right) t^{n-S*_k/2},
\eeq
where the index $k$ label the various classically degenerate twist three trajectories which starts at spin $S^*_k$. This definition of $f$ applied to twist $2$ or for the lowest familty of twist three would lead precisely to $f=I_2$ and $f=I_3$, the two examples considered above. Suppose that $f_S \sim e^{\alpha S} S^\beta$ at large $S$. Then the series converges for $t e^\alpha<1$. Experimentally we observed~$\alpha =0$.\footnote{ Note that the factor of $\Gamma$ functions captures the large spin limit of the sum over structure constants $\sum_k C^2_{S,k}$ that contributes in four-point scalar correlators, and can be determined from, e.g., light-cone bootstrap. That $\alpha=0$ is thus the statement that the individual trajectories share the same asymptotic.}


We claim that $F(t)$ is polynomially bounded with a single cut at $|t|>1$. This in turns implies that $f_{S,k}$ are analytic in spin, which can be made manifest by writing the dispersive representation
\beq
f_{S,k} = \int_1^\infty dt \frac{\text{disc} F(t) }{t^{\tfrac{1}{2}(S-S^*_k)+1}}, \la{fanalytic}
\eeq  
where we drop the contour at infinity (and obtain a convergent expression) for large enough $S$ due to polynomial boundedness of $F(t)$. This can be evaluated numerically following the four step method just outlined.

Combining the analyticity of $f_{S,k}$ with the analyticity of $C_{S,k}^2$ lead to a number of interesting observations:
\begin{itemize}
\item $C_{S,k}$ is analytic in the right half plane (RHP), which in turn implies that $f_{S,k}$ can never vanish in the RHP.
\item $f_{S,k}$ is analytic for $S>S_*$ if and only if $C_{S,k}^2$ has definite sign for $S>S_*$. 
\item Indeed, in \cite{Homrich:2022mmd} we observed that $C_{S,k}^2$ is positive for $S>S_k^*$. In fact, the only zeroes in the RHP are the decoupling zeroes at the even integers smaller than $S_k^*$. This in turn is equivalent to
\beq
\text{disc}F(t) = \sum_{n=1}^{S^*_{k}/2}\frac{a_n}{t^{n}} + o(t^{-S^*_{k}/2}). \la{asymp}
\eeq
Can (a refined version of) this be proved in general CFT?
\item From our new investigations we know that $C_{S,k}^2$ has branch points with ramification index $2$ at points $S_{b_i}$, $b_i$ standing for branch point $i$. The question arises: how this can be generated from (\ref{fanalytic})? These are obtained from specific logarithmic decay of $\text{disc}F(t)$. Specifically, the asymptotic terms
\beq
\text{disc}F(t) \sim t^{(S_{b_i}-S^*_k)/2}(\log t)^{-m/2} \label{cutsSpectrum}
\eeq
with $m\in \mathbb{Z}$ produces a branch point at $S=S_{b_i}$ with ramification index $2$ upon integration. 
\end{itemize}
In sum: From purely integer spin data we can construct a function $F(t)$; with clever coordinates we can explore $F(t)$ beyond its naive convergence domain given by its Taylor series. Moreover, the asymptotic behavior of such a function at large $t$ contains formidable information! \textit{In principle} information about its various cuts can even be extracted as in (\ref{cutsSpectrum}). This can thus \textit{in principle} be used an alternative method to rederive the Riemann surfaces of the previous section as long as we have a monumental amount of integer data to play with as well as ingenious fit methods to extract such asymptotic behaviors from well constructed extrapolations. Does it work \textit{in practice}? See footnote \ref{noWay}.  See also footnote \ref{footnotefits}.\footnote{\label{footnotefits}Although we can't actually use this technique to reconstruct the branch points, we did manage to reconstruct $\text{discF}(t)$ numerically for the first few trajectories and reproduced the asymptotic expansion (\ref{asymp}) with precise determination of the coefficients $a_n$ through fitting of $\text{discF}(t)$. The coefficients $a_n$ control the derivative of the structure constants at the decoupling zeroes. We successfully compared the $a_n$'s determined from $\text{discF}(t)$ with the slopes at the zeroes obtained from the light-ray construction, see e.g.\ figure \ref{csfigure}, or, equivalently, from Newton's series  \cite{Homrich:2022mmd}.}

Hopefully we can learn better spectroscopy methods to make this approach more competitive. 

\section{CFT cousin trajectories}
\label{sec_appendix_newtonconvergence}

The black trajectories in figure \ref{CFTdatafig} provide an analytic continuation in spin of the even-spin twist-three spectrum. They are realized by the light-ray operators constructed in section \ref{lightrayconstruction}, and naturally appear in the Regge limit of four point correlators. 

It is natural to ask whether these analytic continuations are unique. Namely, is there a different way of organizing the spectrum into analytic trajectories? This was one of the questions posed in the introduction. In section \ref{lightrayconstruction}, we argued that the black trajectories correspond to light-ray operators of the form (\ref{ansatzmultitwist}) with sufficiently smooth kernels at coincident points, or, equivalently, light-ray operators (\ref{ansatzmultitwist}) that diagonalize $\mathbb{D}_1$. Nevertheless, this does not exclude the possibility of other analytic continuations existing, even if they are not realized by light-rays of the form  (\ref{ansatzmultitwist}). 

The Newton series (\ref{newtonseries}) provides a numerical strategy to explore this question. Under the assumption of analyticity in the right half $S$ plane and sufficiently tame asymptotic growth in this regime \cite{Buck1948, zbMATH02586461}, e.g.\ exponential type less than $\log(2)$, Newton series converges in the right half-plane. One can then reverse the logic: plugging a selection of even integer values in the right-half plane one can test for convergence of (\ref{newtonseries}). This is how the black trajectories, and the decoupling zeroes of the structure constants of higher trajectories, were originally identified in \cite{Homrich:2022mmd}.  

In this section we \begin{itemize}
    \item Explore this strategy to argue that there exist alternative analytic continuations of the CFT spectrum. 
    \item Identify how to define these trajectories from integrability, at the level of the quantum numbers. 
    \item Speculate on their operator realization and list some open questions.
\end{itemize}
We will omit the operators on the $q=0$ lowest trajectory throughout figures in this section. We should think of them as living in a different symmetry sector as discussed in footnote \ref{lowesttfoonote}.

\begin{figure}[t!]
\centering
\includegraphics[width=0.99\textwidth]{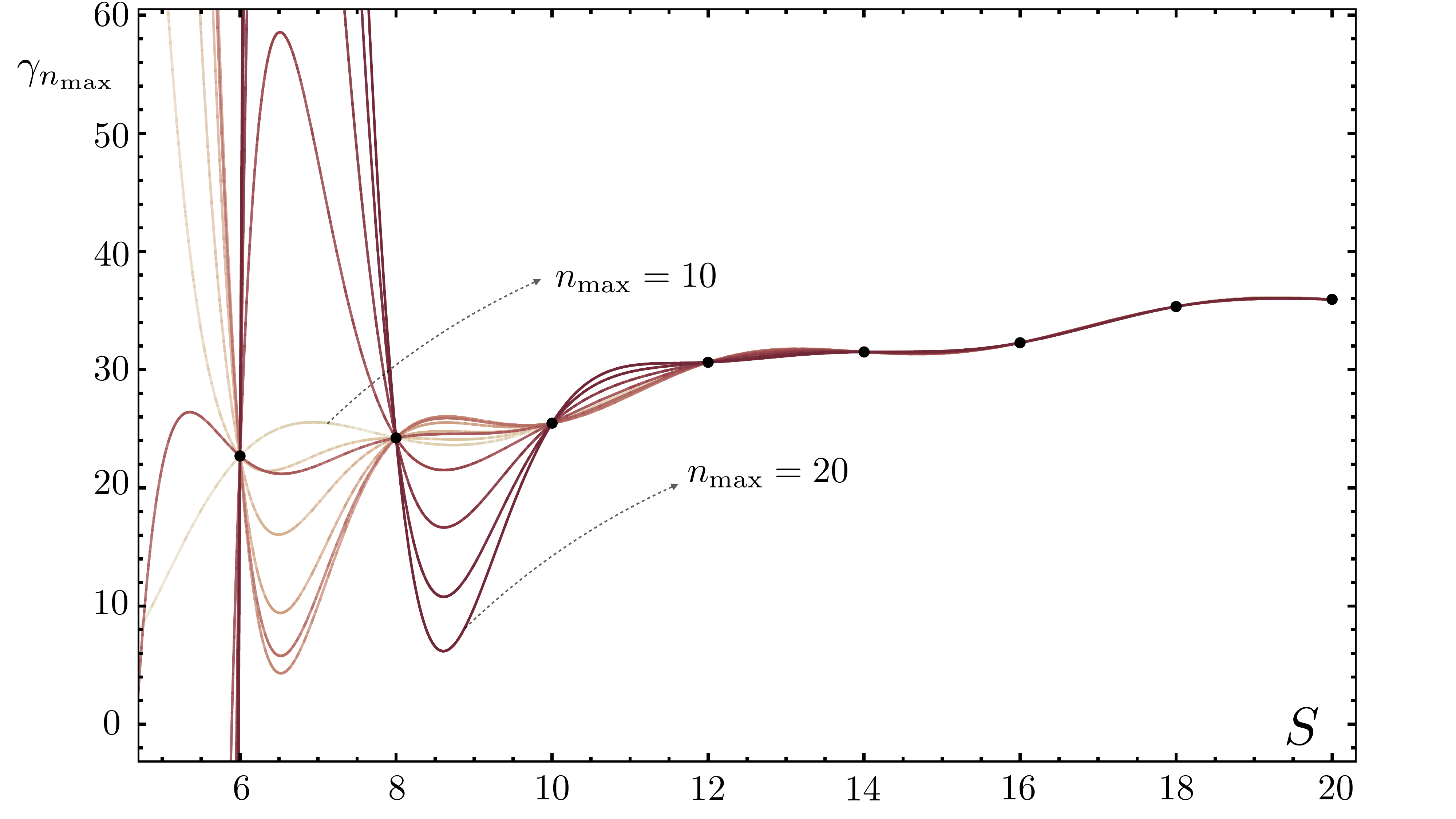}
\caption{The partial sums (\ref{newtonseriesappendixformula}) defining Newton's series constructed with the integer spin red points in figure \ref{CFTdatafig} - the highest anomalous dimension operators at every even spin. From light-colored to dark red $n_\text{max}$ ranges from ten to twenty. The partial sums diverge as more integer points to the right are included in the interpolation. This is expected since there should not be an analytic trajectory satisfying good Regge behavior and analytic in the right half-plane which interpolates the red operators. A similar result would be obtained by applying Newton's series to, e.g., $f(S) = 1/((S-10)^2 + 1)$ or $f(S) = e^{\log(3) S}$.}
\label{redconvergence}
\end{figure}
\begin{figure}[t!]
\centering
\includegraphics[width=0.99\textwidth]{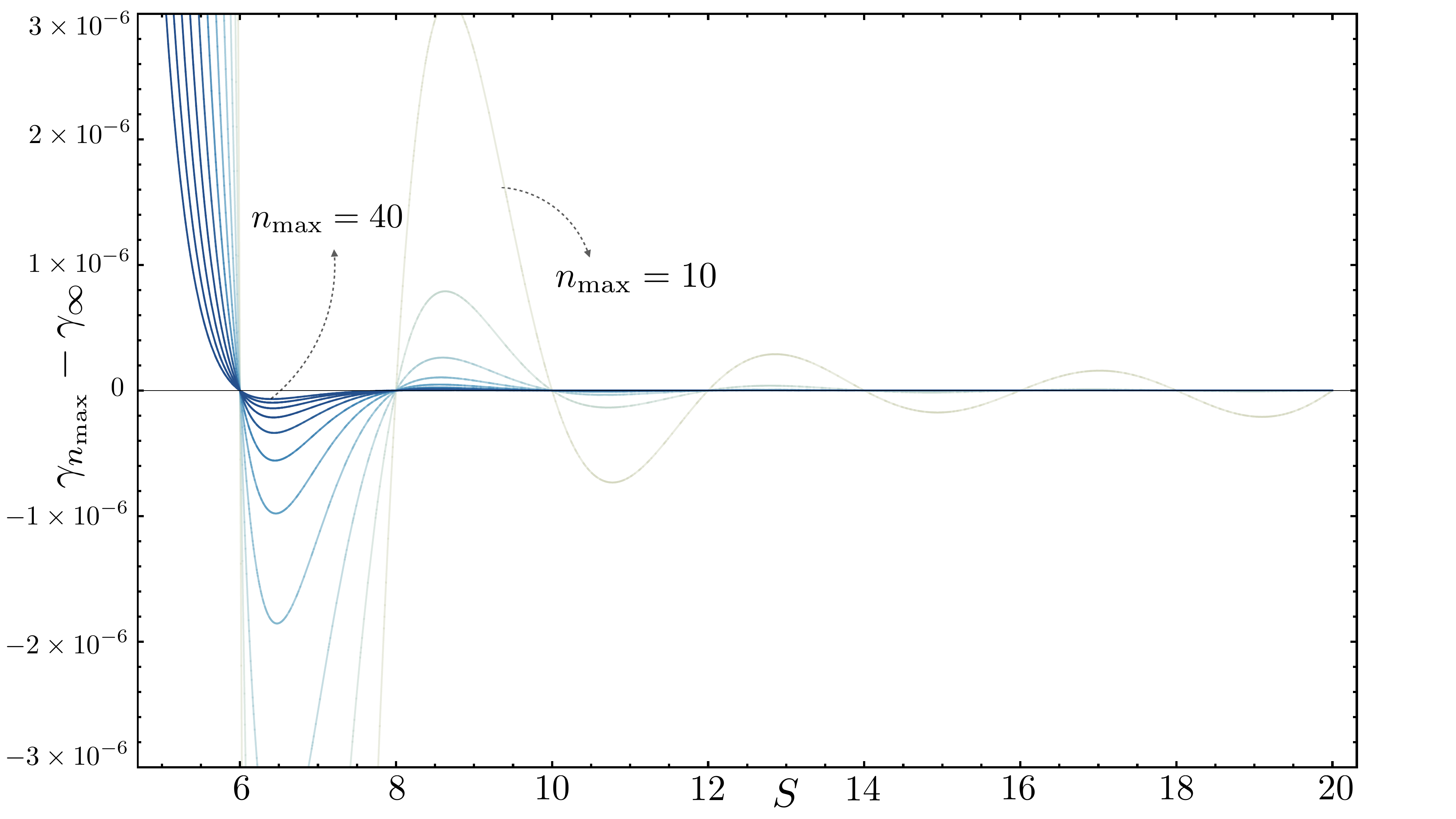}
\caption{For the black trajectories of figure \ref{CFTdatafig} the 
Newton partial sums converge everywhere in the right half-plane. From light-colored to navy $n_\text{max}$ ranges from ten to fourty in steps of four. We plot the error between the partial sums and the `exact' result computed from the Baxter function.}
\label{blueconvergence}
\end{figure}
\begin{figure}[t!]
\centering
\includegraphics[width=0.99\textwidth]{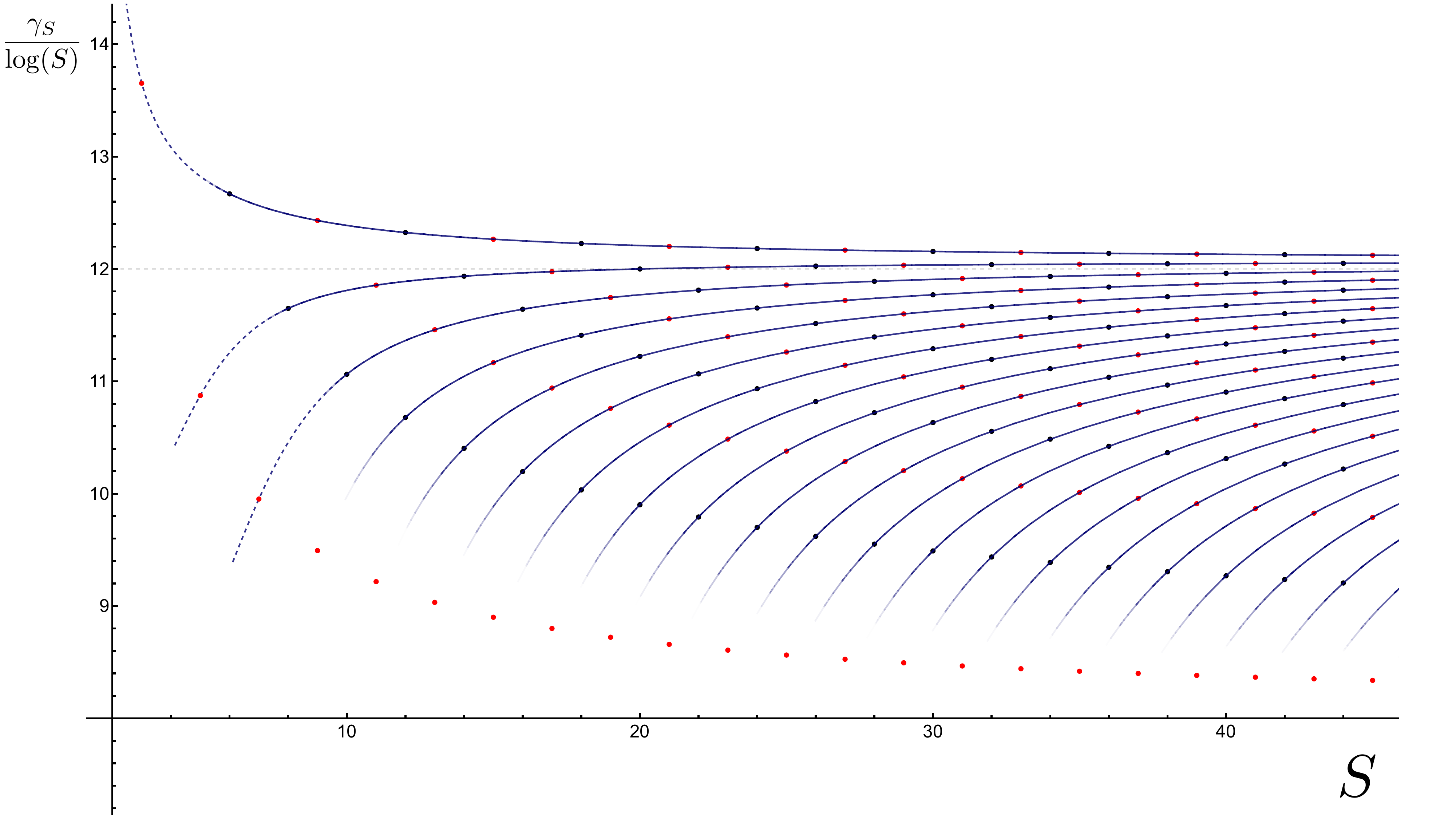}
\caption{Twist three single-trace operators in $\mathcal{N}=4$ SYM can be organized into analytic trajectories in infinitely many ways. Here we organize the spectrum by assigning operators from top to bottom of the spectrum into families. They interpolate physical operators every three units of spin (and thus mix even and odd spin operators). The solid curves in the figure are produced using Newton's interpolation using only the even spin (black points) operators --- that they interpolate the odd spin (red points) is automatic. We fade these curves to the left as they start to lose precision. The dashed curves are produced using Baxter equation as explained in the main body of this section. They perfectly agree with Newton's interpolation. For the first three trajectories we continue them to the left, where Newton's partial sum is not perfectly accurate numerically. More generally, the Baxter equation can be used to construct these trajectories in the full complex spin plane as in section \ref{riemannsurfacesec}.}
\label{figstepsof6}
\end{figure}
\begin{figure}[t!]
\centering
\includegraphics[width=0.99\textwidth]{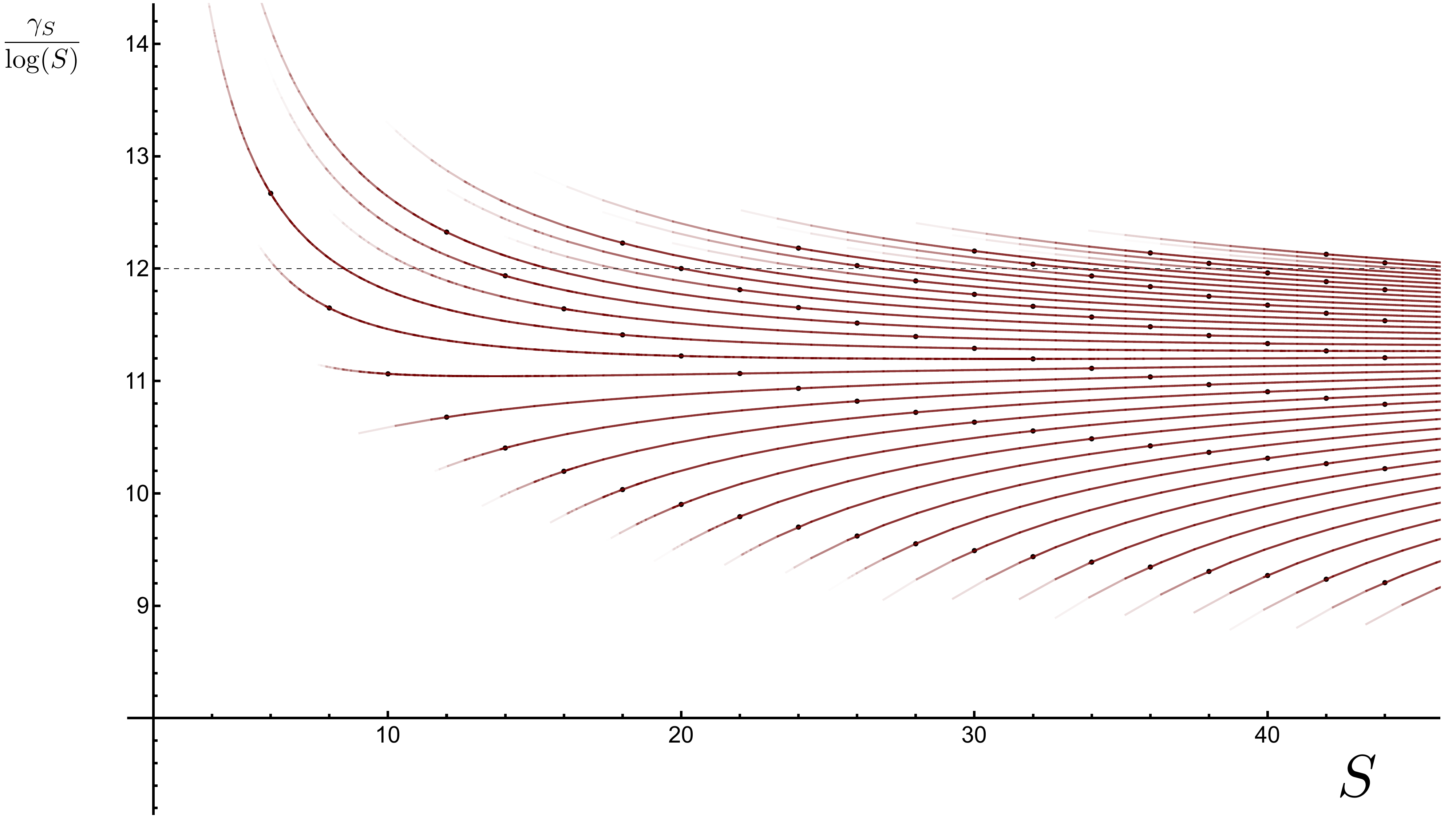}
\caption{The even spin $n=4$ trajectories. They interpolate operators in spin steps of $12$. It is not clear if they are bounded from above or below.}
\label{figstepsof12}
\end{figure}

Let us first illustrate that not anything goes. For example, consider the assignment of operators defined by the red curve in figure \ref{CFTdatafig}. In figure \ref{redconvergence}.(a) we plot the partial sums defined by Newton polynomials
\beq
\gamma_{n_{\text{max}}}(S) = \sum_{n=0}^{n_\text{max}} \binom{\tfrac{S-6}{2}}{n} \sum_{i=0}^n (-1)^{n-i} \binom{n}{i}\gamma^\text{red}_{6 + 2 i}, \label{newtonseriesappendixformula}
\eeq
where we use the first $n_\text{max} + 1$ even operators along the red curve. Compare this with figure \ref{blueconvergence}, where we display the analogous exercise for the second lowest black trajectory of figure \ref{CFTdatafig}. We interpret the result of this exercise as indication that the red list of operators do not form an analytic trajectory (holomorphic on a half plane, with good Regge behaviour at large $S$), while the black list does (as derived from Baxter or from the light-ray construction).

We do, however, identify an infinite number of ways of assigning operators into analytic trajectories! They are peculiar in that these trajectories interpolate the spectrum not in steps of two, but in steps of multiples of three! 

Before describing these trajectories in full generality, let us discuss the first non-trivial example, considered in figure \ref{figstepsof6}. These families interpolate operators in steps of three. The various trajectories are indexed by an integer $\ell\geq0$ and the operators in each trajectory are indexed by $k\geq 1$. Trajectory $\ell$ interpolates the $(1+\lfloor\ell/3\rfloor)$-th operator with highest dimension at spin $3k + 2\ell$. Note that the number of operators at each spin increases in steps of six, hence the origin of the $\lfloor\ell/3\rfloor$ factor, see appendix \ref{countingoperators}. Note that operators are organized from top to bottom, and hence it is not clear if the trajectories are bounded from below. Moreover, these trajectories mix even and odd spin data under analytic continuation!

These trajectories were in practice identified through Newton's series. Plugging the selected local operators in Newton's formula results in beautiful numerical convergence which we interpret as indication that the CFT data indeed extends into an analytic function satisfying the basic properties required for Newton's series convergence. Indeed, note that one could reconstruct this analytic trajectory using only the even spin data along the trajectories using 
\beq
\gamma^\text{steps of 3}_{\ell}(S) = \sum_{n=0}^{\infty} \binom{\tfrac{S-6 -2 \ell}{6}}{n} \sum_{i=0}^n (-1)^{n-i} \binom{n}{i}\gamma^{(1+\lfloor\ell/3\rfloor)\text{-th operator}}_{6 +2 \ell + 6 i}. \label{newtonevenstepsof6}
\eeq
One can then \textit{recover} the odd spin spectrum from (\ref{newtonevenstepsof6}). One obtains a perfect match with the direct computation of the odd spin spectrum. Indeed, the solid curves in figure \ref{figstepsof6} were generated using only even spin data through (\ref{newtonevenstepsof6}). That the red points are interpolated comes automatically. 

Although we identified these trajectories with Newton's method, we understand how to construct them directly at complex spin using the Baxter equation. They are described by solutions of the Baxter  equation (\ref{Baxter}) for entire functions with asymptotics $e^{2\pi u} u^{-2-S}$ as in section \ref{riemannsurfacesec}, but we must drop the zero-momenta condition $Q(i/2) = Q(-i/2)$ in favor of the deformed condition $Q(i/2) = e^{\frac{2 (S - 2l) i \pi }{3}}Q(-i/2)$! We thank Simon Caron-Huot for this suggestion. Imposing this condition fixes $q$ precisely as predicted by the Newton series (which can be applied directly for the $q$ charges). The anomalous dimension extracted from the Baxter function perfectly matches the solid curves in figure \ref{figstepsof6}. Indeed, the dashed curves in figure \ref{figstepsof6} are generated with Baxter equation and extend the curves beyond the regime in which we have strong numerical control with Newton's series. The Baxter equation thus opens the way to explore the full complex plane structure of these new ``steps of three'' trajectories. We anticipate that these various trajectories are connected by branch cuts and form a unique Riemann surface as in section \ref{riemannsurfacesec}, only that now even and odd spin operators are on the same surface.

Infinitely many other assignments are possible. These assignments can be characterized by their ``spin step'' $3n$ with integer $n \geq 3$. They interpolate operators of spin $S = 3n k + S_0$ for integer $k\geq0$. If a given trajectory interpolates the $m$-th highest dimension operator at spin $S$, then it interpolates the $(m+1)$-th highest dimension operator at spin $S + 3n$. We illustrate the assignment $n=4$ in figure \ref{figstepsof12}. Baxter functions associated with these continuations can be constructed through the quantization condition $Q(i/2) = e^{\frac{2 (S - S_0) i \pi }{3n}}Q(-i/2)$.

\section{Discussion} \label{discussionsec}

In this work, we explicitly constructed light-ray operators in $\cN=4$ SYM whose anomalous dimensions and matrix elements provide an analytic continuation in spin of multi-twist local operators. The wavefunctions of these light-ray operators satisfy integral and differential equations coming from integrability. These equations can be solved systematically in a kind of discrete Fourier space, allowing us to plot solutions and numerically compute operator norms and matrix elements. The appropriate solutions satisfy falloff conditions in Fourier space, which give restrictions on the types of singularities the wavefunctions possess at coincident points. These falloff conditions lead to quantization of the integrable charge $q$, and also to discrete Regge trajectories. With explicit light-ray operators in hand, we can also confirm the picture of \cite{Homrich:2022mmd,Henriksson:2023cnh}: there are an infinite number of Regge trajectories, but only a finite number of them have nonzero matrix elements at each integer spin.

An important ingredient in computing matrix elements of our light-ray operators is the light-ray ``norm'' of \cite{Caron-Huot:2013fea,Henriksson:2023cnh}. Our implementation of this norm involves starting with a time-ordered correlator of $2L$ constituent partons, and then integrating them against light-ray wavefunctions. This clearly gives a pairing between wavefunctions, but it is less clearly a pairing between two separate light-ray operators. A nonperturbative approach to this norm is used in \cite{Henriksson:2023cnh}, where they start with a time-ordered four-point function of local operators and integrate it against a pair of bi-local kernels. These two constructions do not correspond to each other in an obvious way, yet they both seem to give consistent results. Can we formulate a more natural definition of a time-ordered correlator of light-ray operators?

The anomalous dimensions of our complex-spin light-ray operators can alternatively be computed from the Baxter equation, by constructing solutions with the correct holomorphicity conditions and large-$u$ asymptotics. Results from the Baxter equation exactly match those coming from our explicit numerical light-ray wavefunctions. Using the Baxter equation, we can also continue further into the complex plane, revealing a rich Riemann surface that unifies na\"ively different Regge trajectories. It is interesting to ask how this Riemann surface interacts with other trajectories such as BFKL trajectories and shadow trajectories that can intersect the Riemann surface at special values of $S$, presumably through a generalization of \cite{Klabbers:2023zdz,Ekhammar:2024neh} to non-parity symmetric trajectories. It would also be interesting to trace how light-ray wavefunctions change in the vicinity of branch points connecting different Regge trajectories. Furthermore, it would be nice to develop methods for computing light-ray matrix elements using separation of variables/integrability without explicitly constructing the wavefunctions, possibly opening the way to computing these observables away from weak coupling. 

An alternative approach to computing data of twist-three light-ray operators in $\cN=4$ SYM was recently presented in \cite{simonlectures}. A central idea of \cite{simonlectures} is to, instead of working with light-ray operator wavefunctions directly, study correlators of local operators in the presence of a light-ray operator. Such correlators are related by a series of integral transforms to the wavefunctions we compute in this work. The approach of \cite{simonlectures} gives an alternative way to efficiently access anomalous dimensions in the right-half $S$ plane.

Although we have developed a relatively complete picture for the simplest analytic continuation in spin of multi-twist trajectories, we also found a surprise: there exist infinitely many other ways of assembling local operators into analytic ``cousin'' trajectories. These cousin trajectories require changing the quantization condition of the Baxter function, which suggests that they involve modifying cyclicity of the trace. Perhaps we can interpret them as light-ray versions of the ``color twist operators'' of \cite{Cavaglia:2020hdb}. Are these cousin trajectories an accident of planar integrability in $\cN=4$ SYM, a feature of single-trace trajectories in large-$N$ gauge theory, or are they a general property of the multi-twist spectrum in CFT? These questions can be tested in particular examples with Newton's series. One particularly interesting feature of cousin trajectories is that they connect even- and odd-spin operators together into a unified object. Is this also a quirk of $\cN=4$ SYM, or does it reflect something more general?

Cousin trajectories have anomalous dimensions asymptoting to $\gamma_S = 3\Gamma_\text{Cusp}\log S$ at large spin. This is to be contrasted with non-cousin trajectories, whose anomalous dimensions asymptote to  $\gamma_S =  2 \Gamma_\text{Cusp} \log S$. At integer spin, the $3 \Gamma_\text{Cusp} \log S$ behavior can be understood from the local operator wave function, which is dominated by spin equally distributed between the various partons. The energy is dominated by the color flux connecting neighboring partons \cite{Alday:2007mf}. Holographically, they correspond to a string made of three folds which approach the boundary of AdS. The folds are connected by a Y-junction which does not approach the boundary. \cite{Belitsky:2003ys, Kruczenski:2004wg}. By contrast, the $2 \Gamma_\text{Cusp} \log S$ behavior corresponds to local operators whose wave function is dominated by two partons having small relative spin or, holographically, the junction approaching the boundary. It would be interesting to understand these asymptotics from a complex spin perspective.  This story is remarkably similar to that of sister trajectories in string theory, whose large spin semiclassics corresponds to spinning strings with extra folds - and thus increased tension \cite{Gross:1987ar, Basso:2014pla}.

Other natural questions about light-ray operators are also interesting to ask about cousin trajectories. The standard analytic trajectories appear naturally in conformal Regge theory \cite{Costa:2012cb,Caron-Huot:2017vep,Kravchuk:2018htv,Caron-Huot:2020nem}, where they describe the behavior of four-point correlators in the Regge limit. What kinematic limit is described by light-ray operators on cousin trajectories? What is the complete story at higher-twist? What do the light-ray wavefunctions of cousin trajectories look like?  Using Newton's series one can construct the analytic continuation of the integrable charge $q$ associated to the trajectories of figure \ref{figstepsof6}. We learn that, for these trajectories, the quantized values of $q$, at fixed $S$, correspond precisely to the values for which the function $\log(\alpha_2/\alpha_1)$ depicted in figure \ref{connectionfig} diverges! In other words, for the values of $q$ corresponding to these trajectories there exist maximally asymmetric solutions of the recursion relation, with Fourier modes asymptotics $a_n \simeq n^{-1}$ for $n\rightarrow \infty$ while $a_n \simeq n^{S}$ for $n\rightarrow -\infty$.\footnote{Recall that operators come in degenerate pairs related by $q \leftrightarrow -q$. This restores the $n \leftrightarrow -n$ symmetry.} It would be interesting to explore this further.

We opened this work reminiscing on the stiff analytic structure imposed on the spectrum of CFTs by large boost physics.  We conclude by noting that, at least in $\mathcal{N}=4$ SYM, likely in holographic CFTs, and ambitiously in general, the story goes much further. Not only operators must organize in analytic families, they must do so in infinitely many ways. How far can we push this incredibly rigid constrain when bootstrapping this and other theories? And what other secrets about the structure of CFTs does Lorentizan physics still hold?

\section*{Acknowledgements}

We thank Carlos Bercini, Simon Caron-Huot, Cyuan-Han Chang, Davide Gaiotto, Kolya Gromov, David Gross, Petr Kravchuk, Martin Kruczenski, Harish Murali, Ian Moult, Enrico Olivucci and Paul Ryan for 
helpful discussions. Research at the Perimeter Institute is supported in part by the 
Government of Canada through NSERC and by the Province of Ontario through MRI.  This work was support in part by the ICTP-SAIFR FAPESP grant 2016/01343-7 and FAPESP grant 2017/03303-1. 
This material is based upon work supported by the U.S.\ Department of Energy, Office of Science, Office of High Energy
Physics, under Award Number DE-SC0011632. This research was supported in part by grant NSF PHY-2309135 to the Kavli Institute for Theoretical Physics (KITP). This work 
was additionally supported by grants from the Simons Foundation (Simons Collaboration on 
the Nonperturbative Bootstrap (DSD: \#488657, PV: \#488661) and Simons Collaboration on 
Confinement and QCD Strings (AH:\#994312)).

\appendix

\section{Counting higher-twist trajectories}

In this appendix, we provide the counting of primary operators at a given spin $S$. The first part of the analysis is performed for general $L$. We then focus on $L=3$. 

For each $L$, to compute the number of single-trace primary states at spin $S$ it is enough to compute the dimension of the space of single-traces of spin $S$, $n_L(S)$. The number of primaries at spin $S$ is then given by $n_L(S)-n_L(S-1)$. We therefore focus on the computation of $n_L(S)$.

The problem of determining $n_L(S)$ is equivalent to the problem of determining the number of permutations of a set of $S+L$ elements, $S$ of which and $L$ of which are indistinguishable, modulo cyclic permutations. In simpler terms, we would like to make a ``necklace'' using ``beads'' of two ``colors'', $L$ of which being blue and $S$ of which being red. We aim to count how many necklaces can be made. Before solving this problem, let us consider a simpler but instructive problem. 

\textbf{Problem 1}
Let us relax the condition that we have $L$ and $S$ beads of each color respectively. Let us simply build necklaces of length $K$ using $m$ colors without tracking how many of each beads of each colors were used.

Consider the set $X$ of all sequences of $K$ beads. We would like to identify those that are equivalent under the action of the cyclic group $G=\mathbb{Z}/K \mathbb{Z}$ and count the number of distinct equivalent classes. In other words, we want to count the number of orbits of the action of $G$ on $X$, $|X/G|$. The result is provided by Burnside Lemma. 

\textbf{Burnside Lemma}
Suppose one is given a set $X$ and an action of a finite group $G$ onto $X$. Denote by $|X^g|$ the number of elements in $X$ that are left invariant by an element $g \in G$. The number of orbits of the action of $G$ onto $X$, $|X/G|$ is given by the average number of fixed points of the action. That is:
\beq
|X/G| = \frac{1}{|G|}\sum_{g \in G} |X^g|.
\eeq
Proof: let $f(g,x) = 1$ if $g.x = x$ and zero otherwise. Then
\begin{align}
\frac{1}{|G|}\sum_{g \in G} |X^g| &= \frac{1}{|G|}\sum_{g \in G} \sum_{x \in X} f(g,x)\\
&= \frac{1}{|G|} \sum_{x \in X} \sum_{g \in G} f(g,x)\\
&= \frac{1}{|G|} \sum_{x \in X} |\text{Stab}(x)|,
\end{align}
where $ |\text{Stab}(x)| $ is the number of elements $g$ that stabilize $x$. But $|\text{Stab}(x)|  = |G|/|O_x|$ where $|O_x|$ is the number of elements in the orbit of $x$ (this is the orbit-stabilizer theorem). We then get 
\begin{align}
\frac{1}{|G|}\sum_{g \in G} |X^g| &=\frac{1}{|G|} \sum_{x \in X} \frac{|G|}{|O_x|} \\
& =\sum_{x \in X} \frac{1}{|O_x|} = |X/G|,
\end{align}
since each orbit $O$ contributes $\frac{1}{|O|}$ $|O|$ times. 

\textbf{Solution of problem 1}

Given a cyclic permutation $g \in G$, how many necklaces are fixed by the action of $g$? A necklace is fixed by the action of $g$ if and only if each cycle of $g$ has a definite color. Suppose that $g$ has $c(g)$ cycles. Then
\beq
|X^g| = m^{c(g)},
\eeq
where $m$ is the number of colours available to build the necklace. 

An element $g \in \mathbb{Z}/K\mathbb{Z}$ has $d$ cycles of length $K/d$, where $d= gcd(K,g)$, the greatest common divisor of $K$ and $g$.   The number of possible necklaces of length $K$ is then
\begin{align}
|X/G|&=\frac{1}{K} \sum_{g =1}^{K} m^{gcd(g,K)}\\
&=\frac{1}{K} \sum_{d | K} \phi\left(\tfrac{K}{d}\right) m^d,  \la{sol1}
\end{align}
where the sum in the second equality is over the divisors of $K$. $\phi(x)$ is the Euler (totient) $ \phi$ function, which counts the numbers of positive integers $y$ smaller than $x$ that are relative prime to $x$ (i.e.\ $gcd(y,x)=1$). The last equality is true because if $gcd(g,K) = d$ then $d | K$, $d | g$ and $gcd(K/d,m/d) = 1$\footnote{Otherwise if $gcd(K/d,m/d) = k $ then $k d$ is a divisor of $g$ and $K$ contradicting that $d$ is the $gcd$.}.

\textbf{Back to counting of primaries}

Equation (\ref{sol1}) provides the solution to problem 1. We can think of (\ref{sol1}) as a partition function. Each quantum number $d$ has degeneracy $\phi(K/d)$, and each state contributes with the number of possible colorings $m^d$. Our original problem requires us to restrict the allowed colorings by requiring that out of the $K = L + S$ beads, $L$ of them are red and $S$ of them are blue. We introduce chemical potential $r$ and $b$ associated to red and blue beads  respectively. The partition (generating) function then becomes

\beq
Z_{L,S}(r,b) =\frac{1}{K} \sum_{d | K} \phi\left(\tfrac{K}{d}\right) \left(r^{\tfrac{K}{d}} + b^{\tfrac{K}{d}}\right)^d . \la{sol2}
\eeq

We interpret each term in the sum as assigning color $r$ or $b$ to each of the cycles $d$ of the original element $g$. Each cycle has $\frac{K}{d}$ elements which must have the same colour. The number of single traces states $n_{L}(S)$ is then simply the coefficient of $r^L b^S$ in the polynomial (\ref{sol2}).

\begin{figure}[t!]
\centering
\includegraphics[width=\textwidth]{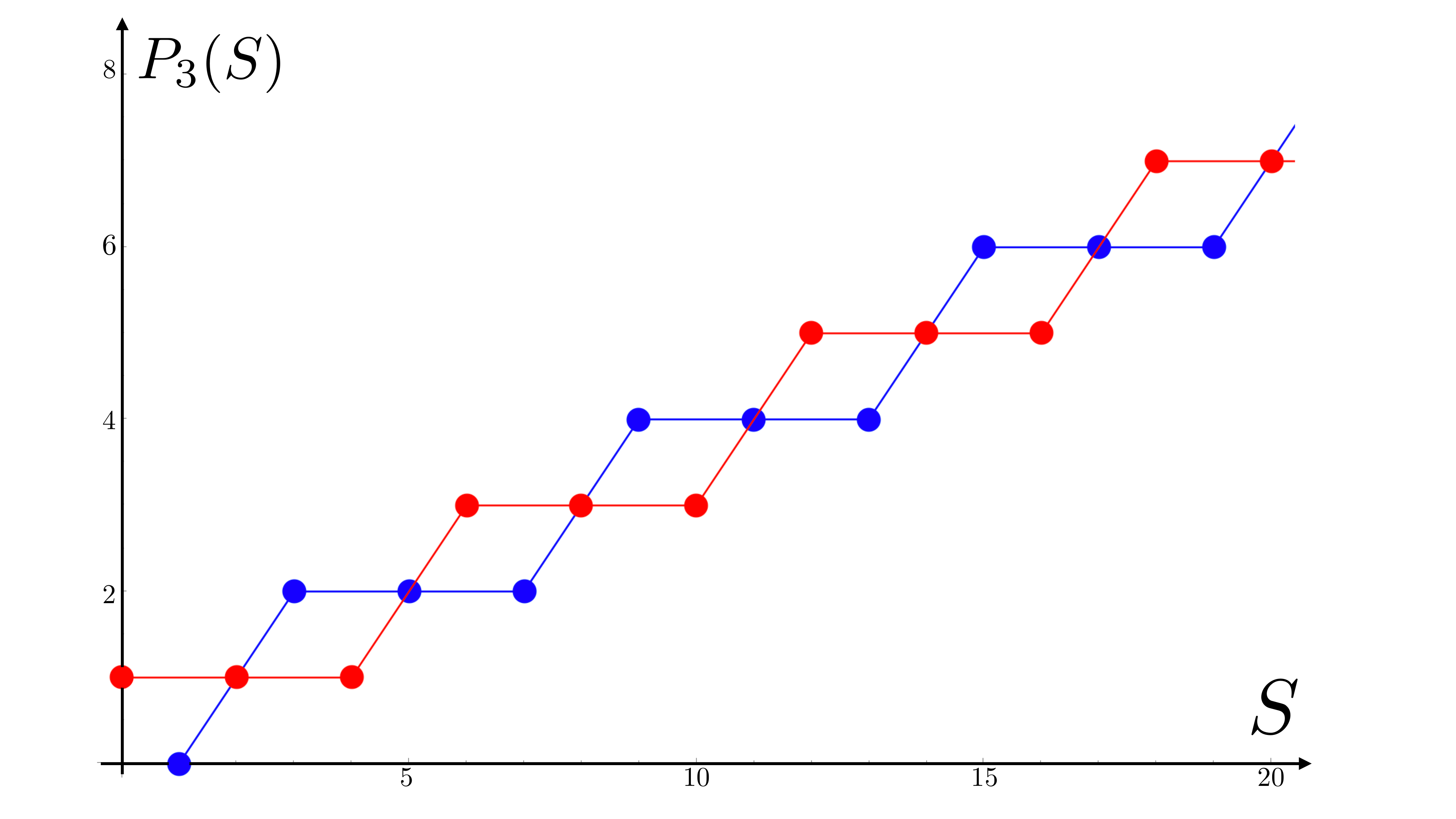}
\caption{Number of twist three primaries as a function of spin. Blue: odd trajectories. Red: even trajectories. They are given by expression (\ref{resultcounting}).}
\label{fig:counting}
\end{figure}

\textbf{Twist three primaries}

Let us set $L=3$ and ask for the coefficient of $b^3$ in (\ref{sol2}). We can set $r=1$ without loss of information. There are only two terms in the sum that can contribute, $d=K$ or $d = K/3$. The latter only contributes if $S$ is a multiple of three. Using that $\phi(1) = 1$ and $\phi(3)=2$, we obtain that 
\beq
n_3(S) =
\left\{
	\begin{array}{ll}
		(S+1)(S+2)/6  & \mbox{if } S/3 \notin \mathbb{Z}, \\
		(S+3)S/6 + 1  & \mbox{if } S/3 \in \mathbb{Z}.
	\end{array}
\right.
\eeq
Note that both answers are integers, as one can check with basic modular arithmetic. The number of primaries $p_3(S) = n_3(S) - n_3(S-1)$ at a given spin $S$ is then
\beq
p_3(S) =
\left\{
	\begin{array}{ll}
		(S+1)/3  & \mbox{if } S \equiv 1 \mod 3,  \\
		(S-1)/3  & \mbox{if }  S \equiv 2 \mod 3, \\
		1+S/3 & \mbox{if }  S \equiv 0 \mod 3.
	\end{array}
\right.
\eeq
To recognize the familiar functions quoted in the introduction, let us consider values of $S \mod 6$:
\beq
p_3(S) =
\left\{
	\begin{array}{ll}
		(S-1)/3  = 2\lfloor S/6\rfloor = 2\lfloor (S+4)/6\rfloor   & \mbox{if } S \equiv 1 \mod 6,  \\
		(S-1)/3 =  2\lfloor S/6\rfloor + 1  & \mbox{if }  S \equiv 4 \mod 6, \\
		(S+1)/3 = 2\lfloor S/6\rfloor + 1  & \mbox{if } S \equiv 2 \mod 6,  \\
		(S+1)/3 = 2\lfloor S/6\rfloor + 2 = 2\lfloor (S+4)/6\rfloor & \mbox{if }  S \equiv 5 \mod 6, \\
		S/3 +1  = 2\lfloor S/6\rfloor + 2 = 2\lfloor (S+4)/6 \rfloor & \mbox{if } S \equiv 3 \mod 6,  \\
	        S/3 +1  = 2\lfloor S/6\rfloor + 1 & \mbox{if }  S \equiv 6 \mod 6. \\
	\end{array}
\right.
\eeq
We therefore conclude that 
\beq
p_3(S) =
\left\{
	\begin{array}{ll}
		2\lfloor (S+4)/6\rfloor   & \mbox{if } S \in 2\mathbb{Z}+1,  \\
		2\lfloor S/6\rfloor + 1  & \mbox{if }  S  \in 2\mathbb{Z},
	\end{array} \la{resultcounting}
\right.
\eeq
as promised. See figure \ref{fig:counting}.

\label{countingoperators}

\section{Details of perturbative computations}
\label{perturbativeappendix}
\subsection{Twist 2}

In this subsection we expand on the computation of the twist two light-ray correlators  $\langle \mathcal{O}_1 \mathbb{O} \mathcal{O}_2 \rangle$  and $\langle \mathcal{T}\{\mathbb{O}\bar{\mathbb{O}} \}\rangle_\text{g.f.} $.

 To compute $\langle \mathcal{O}_1 \mathbb{O} \mathcal{O}_2\rangle$ let us set $y= 0$ and $z = n^+$ in (\ref{twist2lightray}). We take $x_1^- >0$ and $x_2^+<0$. There are two possible Wick contractions. They are related by $(\alpha_1 \leftrightarrow \alpha_2)$, which is a symmetry of the wave function, and therefore amounts to a factor of two. In the end, the computation results in
 \begin{equation}
\langle \mathcal{O}_1(x_1) \mathbb{O} \mathcal{O}_2(x_2)\rangle = \frac{i \sqrt{2}\sqrt{\Gamma(2S+1)}}{2 \pi \Gamma(S+1) x^2_{12}} \int d\alpha_1 d\alpha_2  \frac{\left(\alpha_1 - \alpha_2 + i \epsilon\right)^{-1-S}}{\left(\alpha_1 x_1^- + x_1^2 + i \epsilon\right)\left(\alpha_2 x_2^- + x_2^2 + i \epsilon\right)},
 \end{equation}
 where the second factor in the wave function vanishes by contour deformation. Both integrals are performed by residues, deforming the contour away from the branch cut in the imaginary direction. The final result is
 \beq
\langle \mathcal{O}_1(x_1) \mathbb{O} \mathcal{O}_2(x_2)\rangle = \left(2\pi i \frac{\Gamma(2S+1)}{\Gamma(S+1)^2} \right)  C_{\mathcal{O}_S O_1 O_2}(S)  \frac{\left(\tfrac{x_1^2}{x_1^-} - \tfrac{x_2^2}{x_2^-}\right)}{x_1^- x_2^- x_{12}^2}^{-1-S},\label{final3pt}
 \eeq
which is quoted in (\ref{final3ptmaintext}).
 
 Next, we want to compute (\ref{startingpointtwopoint}) which we reproduce here:
\begin{equation}
\langle \mathcal{T}\{\mathbb{O}\bar{\mathbb{O}} \} \rangle_{\text{g.f.}} = \frac{\Gamma(2S+1)}{4 \pi^2 \Gamma(S+1)^2}\int 
\prod_{i=1}^2d\alpha_i d\beta_i 
\frac{\lambda \delta(\alpha_1 + \alpha_2 -\lambda) \psi(\alpha_1, \alpha_2) \psi(\beta_1, \beta_2)}{(\alpha_1 \beta_1 + d^2 + i \epsilon)(\alpha_2 \beta_2 + d^2 + i \epsilon)}.\label{startingpointtwopoint2}
\end{equation}

Before proceeding with the direct computation, let us emphasize a couple of nice properties of (\ref{startingpointtwopoint2}): 
\begin{itemize}
    \item 
The integrals over $\beta_i$ vanish unless $\alpha_1$ and $\alpha_2$ have opposite sign. To see this, suppose they are positive, and deform $\beta_i$ into the lower half plane, away from the propagator poles. In each term of the wave function, see (\ref{twist2lightray}), only either $\beta_1$ or $\beta_2$ will have singularities in the lower-half plane. One of the integrals thus vanishes, as seen by the contour deformation. 

 \item On the other hand, for even values of $S$, the wave functions have no support when $\alpha_i$ have opposite sign, see discussion under (\ref{doubletwistlightray}). 
\end{itemize}
We thus expect that, for even spin, when the light-ray reduces to $L[\mathcal{O}_S]$, the continuous spin norm defined by ${\langle \mathcal{T}\{\mathbb{O}\bar{\mathbb{O}} \} \rangle_{\text{g.f.}}}/{\langle 0 | \mathbb{O} \bar{\mathbb{O}} |0\rangle}$ vanishes. We thus see that the conversion between the continuous spin norm and the continuation in $S$ of the canonical norm of the local operators $\mathcal{O}_S$ --- which equals one --- must involve dividing by zeroes at every even integer. We will see this explicitly. 

To compute (\ref{startingpointtwopoint2}), let us instead perform the integrals over $\beta_i$ by picking the poles of the propagators. Change variables as $u = \alpha_1 + \alpha_2$, $v = \alpha_1 - \alpha_2$. The integral in $u$ is fixed by the gauge-fixing. Since $\alpha_i$ must have opposite sign, it is enough to integrate over $|v|>\lambda$. In fact, from permutation symmetry of $\psi(\alpha_1,\alpha_2)$ we can restrict to $v>\lambda$ up to a factor of two. We will use the fact that $\psi(\alpha_1, \alpha_2) = \left( 1 + e^{i\pi (-1-S)}\right)|v|^{-1-S}$. The integral reduces to
\beq
 \frac{4 \Gamma(2S+1)}{\Gamma(S+1)^2} \left( 1 + e^{i\pi (-1-S)}\right) \lambda \int_{v>\lambda} dv \frac{v^{-1-S}}{\lambda^2 - v^2} \left(\frac{2d^2}{v+\lambda}+\frac{2d^2}{v-\lambda}\right)^{-1-S} = -\frac{ \left( 1 + e^{i\pi (-1-S)}\right)}{(2S+1) d^{2+S}}  \label{finaltwopt}.
\eeq
The factor in the numerator realizes the manifest zeroes described above. The denominator, in a similar manner to what happened in the three-point matrix element case, can be understood from the (double) light-transform of the canonical two-point structure. Indeed, see \cite{Kravchuk:2018htv}, 
\beq
\langle 0|L[ \mathcal{O}_S] L[\bar{\mathcal{O}}_S] | 0 \rangle =  \frac{-2 \pi i}{\tau + 2S - 1}\langle 0 | \mathbb{O} \bar{\mathbb{O}} |0\rangle = {\frac{-2\pi i}{(2S+1)d^{2+S}}},
\eeq
where $\langle 0 | \mathbb{O} \bar{\mathbb{O}} |0\rangle$ is the canonical two-point structure for operators of dimension $1-S$ and spin $1-\Delta$. In the last equality we specified our coordinate frame and twist two. We thus conclude that 

\beq
\frac{\langle \mathcal{T}\{\mathbb{O}\bar{\mathbb{O}} \} \rangle}{\langle 0 | \mathbb{O} \bar{\mathbb{O}} |0\rangle} = -\left(1+ e^{i \pi (-1-S)}\right)/(2S+1).
\eeq

\subsection{Twist 3}

In this appendix we examine the computation of $\langle \mathcal{O}_1 \mathbb{O}_{n,S} \mathcal{O}_2 \rangle$ and ${\langle \mathcal{T}\{\mathbb{O}_{n,S}\bar{\mathbb{O}}_{n,S} \} \rangle}$. 

We parametrize twist three light-rays $\mathbb{O}_{n,S}$ inserted at $x=-\infty n^+$, $z=n^+$ as
\beq
\mathbb{O}_{n,S} = \int d\alpha_{\text{cm}} dr d\theta \left(\psi\left(\alpha_{cm},r,\theta\right) \equiv \frac{1}{\Gamma(S_n^* -S)}\frac{g_{n,S}(\theta)}{r^{1+S}}\right) \text{Tr}\left(\mathcal{Z}(\alpha_1) \mathcal{Z}(\alpha_2) \mathcal{Z}(\alpha_3)\right),
 \eeq
with $\alpha_i$ being parametrized by $(\alpha_{cm},r,\theta)$ according to (\ref{alphachange1}-\ref{alphachange3}) as 
\begin{align}
\alpha_i &=  \alpha_{cm} + \tfrac{1}{3} r \alpha_i^\theta, \\ 
\alpha_i^\theta &= \cos(\theta - 2\pi (i-1)/3) - \cos(\theta - 2\pi (i-2)/3).
\end{align}
 We introduce the $\Gamma$ function factor for later convenience, where $S_n^* = 6 (n-1)$ is the first integer spin in which the $n$-th even trajectory has physical operators, see figure \ref{CFTdatafig}. 

The computation of the matrix elements $\langle \mathcal{O}_1 \mathbb{O}_{n,S} \mathcal{O}_2 \rangle $ is straightforward. We take $\mathcal{O}_1 = \Tr\left(\mathcal{X}\bar{\mathcal{Z}}\right)$ and $O_2 = \Tr\left(\bar{\mathcal{X}}\bar{\mathcal{Z}}^2\right)$ and consider $x_1^->0$, $x_2^-<0$. All Wick contractions are equivalent due to cyclic symmetry of the wave function, and amount to a factor of six. We thus want to compute
\beq
\langle \mathcal{O}_1 \mathbb{O}_{n,S} \mathcal{O}_2 \rangle  = 6 \int d\alpha_{\text{cm}} dr d\theta \frac{\psi\left(\alpha_{cm},r,\theta\right) (x_2^- \alpha_{3} + x_2^2 + i \epsilon)^{-1}}{ (x_1^- \alpha_{1} + x_1^2 + i \epsilon)(x_2^- \alpha_{2} + x_2^2 + i \epsilon) }.
\eeq

The integral over $\alpha_{cm}$ can be performed by deforming the contour into the lower half-plane, picking up the pole at $\alpha_{cm}/3 =  - x_1^2/x_1^- - r \alpha_1^\theta - i \epsilon$.  The integral over $r$ can also be performed by residues by realizing that distributionally
\beq
\int_{\mathbb{R}^+} dr r^{-1-S} f(r) = \int_\mathbb{R} dr \frac{e^{-i\pi(1+S)} (r - i \epsilon)^{-1-S}-e^{i\pi(1+S)} (r + i \epsilon)^{-1-S}}{2 i \sin(\pi S)} f(r). \label{radial}
\eeq
One can then perform the integral over $r$ in each term by deforming away from the branch-points in the imaginary direction, thus picking the propagator poles.\footnote{Whether each 
propagator pole is in the upper or lower half-plane depends on the value of $\theta$. However, the pole is always picked in either the first or the second term in (\ref{radial}). One thus needs to be careful with the appropriate $i \epsilon$s and phases, which is uninteresting and straightforward. When the dust settles, the various contributions combine into the simple result (\ref{finaltheta}).}  The final result
for $\langle \mathcal{O}_1 \mathbb{O}_{n,S} \mathcal{O}_2 \rangle $ is
\begin{align}
& \frac{12 i \pi^2  3^{-S}}{x_1^-(x_2^-)^2 \sin(\pi S) \Gamma(S^*_n - S)}\left(\frac{x_1^2}{x_1^-} - \frac{x_2^2}{x_2^-}\right)^{-2-S}
 \int  d\theta g(\theta)  \frac{\left({\alpha_3^\theta - \alpha_1^\theta} - i \epsilon \right)^{1+S}-\left({\alpha_2^\theta - \alpha_1^\theta} - i \epsilon \right)^{1+S}}{(\alpha_3^\theta-\alpha_2^\theta) }, \label{finaltheta}
\end{align}
which must be evaluated numerically for each trajectory as we will discuss later. From (\ref{finaltheta}) we read that the kernel of (\ref{maintextfinalthreepoint}) is 
\beq
 F_{\text{three points}}(\theta) = 12 i \pi^2  3^{-S} \times  \frac{\left({\alpha_3^\theta - \alpha_1^\theta} - i \epsilon \right)^{1+S}-\left({\alpha_2^\theta - \alpha_1^\theta} - i \epsilon \right)^{1+S}}{(\alpha_3^\theta-\alpha_2^\theta) }. \label{kernelofmaintext3pt}
\eeq

Next, let us consider the computation of the norm $\langle\mathcal{T}\{\mathbb{O}\bar{\mathbb{O}}\}\rangle_\text{g.f.}$. We will consider the same crosswise kinematical configuration as in the twist-2 case, see figure \ref{twist3comp}. We use the gauge fixing $ \delta(r_\alpha - 1)$. In a similar manner to the previous case, the correlator vanishes if all $\alpha_i$ have the same sign. We can use cyclicity and reflection symmetry of the $\psi(\alpha_i)$ wave function, equations (\ref{rinv}) and (\ref{cinv}), to restrict to the case $\alpha_1>0$, $\alpha_2<0$, $\alpha_3 <0$. All planar Wick contractions can be related by cyclicity of the wave function $\psi(\beta_i)$. In sum, we would like to compute

\beq
\langle\mathcal{T}\{\mathbb{O}\bar{\mathbb{O}}\}\rangle_\text{g.f.} = 18 \int_{\substack{\alpha_1 >0\\\alpha_2<0\\\alpha_3<0}}  \frac{dr_\alpha d\alpha_{cm} d\theta_\alpha  dr_\beta d\beta_{cm} d\theta_\beta \delta(r_\alpha- 1) \psi_{\mathbb{O}}\left(\alpha_{cm},r_\alpha,\theta_\alpha \right)\psi_{\bar{\mathbb{O}}}\left(\beta_{cm},r_\beta,\theta_\beta\right) }{ (\beta_1  \alpha_{1} + d^2 + i \epsilon)(\beta_2 \alpha_{2} + d^2 + i \epsilon) (\beta_3 \alpha_{3} + d^2 + i \epsilon)}. \nonumber
\eeq

The integrals over $\beta_{cm}$ and $r_\beta$ can be performed by picking the propagator poles, as in the three-point matrix element computation. The integral over $r_\alpha$ is localized. We are left with a triple integral
\begin{align}\label{twopointintegral}
&\langle\mathcal{T}\{\mathbb{O}\bar{\mathbb{O}}\}\rangle_\text{g.f.} =  \frac{(d^2)^{-1-S}}{\sin(\pi S) \Gamma(S_n^* - S)^2} \int_D d\alpha_{cm} d\theta_\alpha d\theta_\beta g_{\mathbb{O}}(\theta_\alpha)g_{\bar{\mathbb{O}}}(\theta_\beta) \times \\
& \bunderbrace{\left(\frac{\left(\left((\alpha_{cm} + \alpha_1^\theta)(-\alpha_{cm} - \alpha_2^\theta)\right)^{1+S} F_1(\theta_\alpha, \theta_\beta) - \left((\alpha_{cm} + \alpha_1^\theta)(-\alpha_{cm} - \alpha_3^\theta)\right)^{1+S} F_2(\theta_\alpha, \theta_\beta)  \right)}{(-36 \pi^2 i (3)^{-2S} e^{i\pi (S+1)})^{-1}\left(F_3(\theta_\alpha, \theta_\beta) + \alpha_{cm} F_4(\theta_\alpha, \theta_\beta) \right) } \right)}{ \let\scriptstyle\textstyle F_{\text{two point}}(\theta_\alpha, \theta_\beta,\alpha_{cm})},\nonumber
\end{align}
from which we read the kernel $F_{\text{two point}}$ of (\ref{finalmaintextnorm}) which is defined in terms of
\begin{align*}
F_1(\theta_\alpha, \theta_\beta)  &= \left(\frac{\alpha_1^\theta - \alpha_2^\theta}{\beta_1^\theta - \beta_2^\theta} + i \epsilon \right)^{-1-S}\hspace{-0.9cm}, \qquad F_2(\theta_\alpha, \theta_\beta)  = \left(\frac{\alpha_1^\theta - \alpha_3^\theta}{\beta_1^\theta - \beta_3^\theta} + i \epsilon \right)^{-1-S}\hspace{-0.9cm}, \\F_3(\theta_\alpha, \theta_\beta)  &= \alpha_1^\theta \beta_1^\theta (\alpha_2^\theta - \alpha_3^\theta)+\alpha_2^\theta \beta_2^\theta (\alpha_3^\theta - \alpha_1^\theta) + \alpha_3^\theta \beta_3^\theta (\alpha_1^\theta - \alpha_2^\theta), \\ F_4(\theta_\alpha, \theta_\beta)  &= \beta_1^\theta(\alpha_2^\theta - \alpha_3^\theta) + \beta_2^\theta(\alpha_3^\theta - \alpha_1^\theta) + \beta_3^\theta(\alpha_1^\theta - \alpha_2^\theta) \label{Ftwopointdef}
\end{align*}
and we re-scaled $\alpha_{cm} \rightarrow \alpha_{cm}/3$. The domain of integration $D$ is given by $\alpha_{cm} > -\alpha_1^\theta$, $\alpha_{cm} < -\alpha_2^\theta$, $\alpha_{cm} < -\alpha_3^\theta$. The integral over $\alpha_{cm}$ can be evaluated analytically. We do not use this fact. 

Finally, in order to be able to write (\ref{maintextfinalthreepoint}) and (\ref{finalmaintextnorm}) we define  our choices of two- and three- point structures. In our conformal frames, we choose \beq
    \langle 0 | \mathbb{O} \bar{\mathbb{O}} |0\rangle = (d^{2})^{-1-S} \qquad \text{and} \qquad \langle0|\mathcal{O}_1 \mathbb{O} \mathcal{O}_2|0\rangle = \frac{1}{x_1^-  (x_2^-)^{2}} \left(\frac{x_1^2}{x_1^-} - \frac{x_2^2}{x_2^-}\right)^{-2-S}.
    \eeq

\section{Q-functions from light-rays}
\label{baxterappendix}

In section \ref{lightrayconstruction}, we emphasized that light-ray operators in $\mathcal{N}=4$ SYMs should diagonalize the transfer-matrix operator, whose action on light-rays of arbitrary twist is defined through equation (\ref{taudef}). The transfer matrix operator was very useful due to its simple differential action. Its polynomial eigenvalues are sufficient to completely characterize the state. 

The light-rays should also diagonalize the Baxter $\mathbb{Q}$ operator. It act as an integral operator \cite{Derkachov_1999}
\begin{align}
\nonumber &\mathbb{Q}(u) \circ \mathbb{O}= \frac{\cosh(\pi u)^L}{\pi^L}\int_0^1 \prod_{j=1}^L dz_j (1-z_j)^{i u -1/2}(z_j)^{-i u -1/2} \times  \\& \int_{-\infty}^{\infty} \prod_{i=1}^L d\alpha_i \psi(\alpha_1,\dots,\alpha_L)\text{Tr}\left[\mathcal{Z}(z_1 \alpha_1 + (1-z_1) \alpha_{L}) \dots \mathcal{Z}(z_L \alpha_L + (1-z_L) \alpha_{L -1}) )\right],\la{Qop}
\end{align}
and is especially useful due to serving as a generating function of (infinitely many) integrable charge operators
\beq
\mathbb{C}_k^\pm = 2i^{k+1}\left(\mathbb{Q}^{(k)}(i/2)\mp \mathbb{Q}^{(k)}(-i/2)\right).
\eeq
In particular, the dilatation operator, whose action was defined in (\ref{dilatationdef}), is given by $\mathbb{D} = \mathbb{C}^+_1$. Note that the Baxter operator is an entire function of $u$. The poles of the integral in (\ref{Qop}) at $u =  k \pm i/2$ cancel with the zeroes of the trigonometric prefactor. The prefactor is normalized so that 
\beq
\mathbb{C}^+_0 = 1,
\eeq
which follows from cyclicity of the trace.

The eigenvalues
\beq
\mathbb{Q}(u) \circ \mathbb{O} = Q(u) \mathbb{O}
\eeq
can be obtained without reference to $\psi$. This is the strategy adopted in section (\ref{riemannsurfacesec}) in order to uncover the operator mixing in the spin left half-plane. For that, one uses the operatorial Baxter equation
\beq
\mathbb{Q}(u+i) (u+i/2)^L + \mathbb{Q}(u-i)(u-i/2)^L = \mathbb{Q}(u) \tau(u). \la{operatorialbaxter}
\eeq

Acting on a light-ray operator produces the Baxter equation (\ref{Baxter}) but, crucially, note that the various quantization conditions for complex spin, first proposed by Janik in the twist two context \cite{Janik:2013nqa} and generalized to higher twist in \cite{Homrich:2022mmd}, are here \textit{fixed} given general properties determing the light-ray wave functions. In particular, as argued above, the zero momentum condition
$$Q(i/2)=Q(-i/2)$$ holds at complex spin due to cyclicity of the trace defining light-ray operators,\footnote{A priori, one might have wondered if such a condition should be non-trivially modified for non-integer spin.} the analyticity of $Q(u)$ follows from the analyticity of $\mathbb{Q}$, and, we expect, the leading asymptotics of $Q(u)$ are fixed by the Lorentz spin of the light-ray $J_L$
\beq
\lim_{u\rightarrow  \infty}Q(u) \propto e^{2\pi u}  u^{J_L} + \dots\,,
\label{asympresult}\eeq
where $J_L = 1- L - S$, see section \ref{weakcouplingsec}, thus providing a physical explanation for the decaying solution of Janik. These three properties, combined with the Baxter equation (\ref{Baxter}), uniquely fix all Baxter eigenfunctions corresponding to the twist three trajectories at any value of the spin. 
In section \ref{riemannsurfacesec} we discuss how to do so numerically and present results for twist three. 

We have not been successful in deriving the result (\ref{asympresult}) for general $L$. However, it is interesting to see that, for $L=2$, (\ref{asympresult}) --- and, therefore, Janik's $Q$-function --- can be derived from the light-ray wave function $\psi_2$. To do so, we start from (\ref{Qop}). Changing variables as $(z_i \alpha_i + (1-z_i) \alpha_{i-1}) \rightarrow \alpha_i$ we obtain
\begin{align}
\nonumber &\mathbb{Q}(u) \circ \mathbb{O} =\frac{\cosh(\pi u)^2}{\pi^2}\int_0^1 \prod_{j=1}^2 dz_j (1-z_j)^{i u -1/2}(z_j)^{-i u -1/2} \times  \\& \int_{-\infty}^{\infty} \prod_{i=1}^2 d\alpha_i |1-z_1 -z_2|^{-1} \psi_2\left(\frac{\alpha_i}{-1 + z_1 + z_2}\right) \text{Tr}\left[\mathcal{Z}(\alpha_1)  \mathcal{Z}(\alpha_2) )\right].
\end{align}
Using that $\psi_2(\lambda \alpha_i) = |\lambda|^{J_2 \equiv -S-1}\psi_2(\alpha_i)$ we get
\begin{align}
&\mathbb{Q}(u) \circ \mathbb{O} = \Bigg(Q_2(u) =  \frac{\cosh(\pi u)^2}{\pi^2}\int_0^1 dz_1 g(z_1) \times\\&\left(\int_0^{1-z_1}  dz_2 (1-z_1 -z_2)^{S}g(z_2)+ \int_{1-z_1}^1 dz_2 (-1+z_1 +z_2)^{S} g(z_2) \right) \Bigg) \times \mathbb{O}  \nonumber
\end{align}
with $g(x) =(1-x)^{i u -1/2}(x)^{-i u -1/2}$. At large $u$, the integrals can be performed by saddle point. The saddles are located at $z_1 - z_2 = 0$, $z_1 +z_2 = 1$. The large $u$ behaviour of $Q_2(u) \sim u^{-1-S}$ is then manifest. Of course, the integrals above are of the Euler type and can be performed exactly, matching Janik's Q-function at finite $u$.

\bibliographystyle{JHEP}
\bibliography{refs}

\end{document}